\documentclass[twocolumn]{revtex4}
\usepackage{color}
\usepackage{listings}            
\usepackage{subfigure}           
\begin{document}

\title{Quantization of free electromagnetic field: A different approach to the time-like polarization}
\author{Klaudia Wrzask}
\affiliation{Department of Control and Power Engineering,
	Gda\'nsk University of Technology, Poland}

\begin{abstract}
	A four-dimensional photon polarization space, such that gives a different interpretation of the ladder operators for the time-like degree comparing to the Gupta-Bleuler formulation is presented. This interpretation, coming from the construction of a covariant Hamiltonian, gives positive defined norms for all the four polarization degrees of freedom. This paper is written in the background of reducible representation algebras introduced by Czachor, although the interpretation of the ladder operators as presented here could be done independently. States that reproduce standard electromagnetic fields (i.e. photons with two transverse polarizations) from the four-dimensional covariant formalism (i.e. with two additional longitudinal and time-like polarizations) are shown explicitly. Lastly, Lorentz and gauge transformations are discussed in detail.
\end{abstract}
\maketitle
\section{Introduction}
The Gupta-Bleuler formulation of four-dimensional \linebreak quantization is well known from literature \cite{Gupta50} - \cite{WHK03}. \linebreak Mostly, it is assumed that the time-like photon states have negative norms. Unfortunately, states with such norms form a serious difficulty with the probability interpretation of quantum mechanics. There seems to be no experimental verification of such particles, therefore they are often called in literature unphysical or even ``ghosts" \cite{HPDER69}. In this paper the problem of negative norms of the time-like photons is profoundly investigated, introducing a different interpretation of the ladder operators.

For further analysis let us take a convention $c=\hbar=1$. Building a relativistic model for photons, we have to consider momentum and polarization of the photon. In this mathematical model these two quantities will be described in a tensor product structure, i.e.
\begin{eqnarray}
\textrm{photon momentum space}~
\otimes
~
\textrm{photon polarization space},
\nonumber
\end{eqnarray}
although it should be stressed that in relativistic context spin and momentum are not independent degrees of freedom.

First, in Section \ref{sec:4dconstruction} a construction coming from a covariant Hamiltonian for a four-dimensional polarization oscillator is shown. 
Then we ask what is the consequence of creating particles on the energy of the whole system, i.e. does this raise the energy level or lower it? This is discussed in sections: \ref{sec:4dconstruction123} for the space-like polarizations and further in \ref{sec:4dconstruction0e+} and \ref{sec:4dconstruction0e-} two different interpretations of the time-like polarization are considered. 

Next, some preliminary aspects on the Czachor's reducible representations of canonical commutation relations (CCR) are presented. Such model has strong arguments in its favor, mainly because it deals with most of the infrared and ultraviolet divergences. In Section \ref{sec:redmot} a basic concept of the model is introduced. Further, in Section \ref{sec:redladder} the ladder and the number operators are introduced for $N=1$ oscillator space.
In Section \ref{sec:redHam} the covariant Hamiltonian in the reducible representation algebra is presented. 
How does the extension to $N$-oscillator space look like and what is the definition of the number operator in an arbitrary $N$ oscillator space are discussed in Section \ref{sec:redladderN}. 
Next, the construction of vacuum is shown in \ref{sec:4dvacuum}.  Then we ask: how does such a theory, with four polarization degrees of freedom, correspond to Maxwell's electromagnetism theory. It turns out that vectors, denoted here by $\Psi_{EM}$, reproduce standard Maxwell electrodynamics. This is shown in Section \ref{sec:4dHem}. Finally, the electromagnetic field tensor for such representation is introduced in Section \ref{sec:4dF}. 

In the last section a homogeneous Lorentz transformation for the four-dimensional oscillator is introduced. When taking into account the four-photon polarization degrees of freedom, the Lorentz transformation is accompanied by another transformation and this manifests itself also on the spin-frame level. In Section \ref{sec:LorentzBogoliubov} a corresponding transformation of the ladder operators is derived. Further, in Section \ref{sec:Lorentzrepresentation}, the generators for the irreducible representation are introduced. Next, in Section \ref{sec:Lorentzreducible}, we introduce Lorentz transformation in reducible representation algebra, whereas, in \ref{sec:Lorentzvacuum} we show vacuum transformation. In Section \ref{sec:gaugetransfor} we discuss gauge transformations of the four-vector potential and
in Section \ref{secmixing} we show that the ``ghost" operator and the covariant number operator are invariant in any gauge and any reference frame. Finally, in Section \ref{sec:P} the four-translations are introduced. Summary and final conclusions are drawn in Section \ref{sec:4dredconclusions}. 

This paper closes with mathematical appendices, introducing  the Minkowski and null tetrad in Penrose and Rindler notation \cite{RPWR84} and proofs involving Lorentz transformation on electromagnetic potential electromagnetic field operator.
\section{Four-dimensional photon polarization space}
\subsection{Construction of four-dimensional polarization space}\label{sec:4dconstruction}
To construct a four-dimensional photon polarization space, let us first introduce a covariant Hamiltonian of the form
\begin{equation}
H\label{covH}
=
-
\frac{p_{{\bf a}}p^{{\bf a}}}{2}
-\frac{q_{{\bf a}}q^{{\bf a}}}{2},~~~ {\bf a}=0,1,2,3.
\end{equation}
Here $p_{{\bf a}}$ and  $q_{{\bf a}}$ are some canonical variables such that
\begin{equation}\label{canvar}
p_{{\bf a}}
=
i\partial_{{\bf a}}
=
i\frac{\partial}{\partial q^{{\bf a}}}
=
ig_{{\bf {ab}}}\frac{\partial}{\partial q_{{\bf {b}}}},
\end{equation}
with commutation relations
\begin{eqnarray}
[q_{{\bf a}},p_{{\bf {b}}}]
&=&
-ig_{{\bf{a}}}{_{{\bf{b}}}}.
\end{eqnarray}
The metric here is denoted as diag$(+,-,-,-)$. These canonical variables should not be mistaken with the position and momentum of the photon field. This construction is made strictly for the four degrees of photon polarization.  Now let us define non-hermitian operators
\begin{eqnarray}\label{acov}
a_{{\bf a}}
&=&
\frac{
	q_{{\bf a}}+ip_{{\bf a}}}{\sqrt{2}}
=
\frac{
	q_{{\bf a}}-\partial_{{\bf a}}}{\sqrt{2}},
\\
a_{{\bf a}}^{\dagger}\label{acovdag}
&=&
\frac{
	q_{{\bf a}}-ip_{{\bf a}}}{\sqrt{2}}
=
\frac{
	q_{{\bf a}}+\partial_{{\bf a}}}{\sqrt{2}},
\end{eqnarray}
which satisfy the following commutation relations
\begin{equation}\label{comaadag}
[a_{{\bf a}},a_{{\bf {b}}}^{\dagger}]
=
-g_{{\bf{a}}}{_{{\bf{b}}}},
\end{equation}
\begin{equation}
[a_{{\bf a}},a_{{\bf {b}}}]\label{comaa}
=
[a_{{\bf a}}^{\dag},a_{{\bf {b}}}^{\dag}]
=
0.
\end{equation}
As a consequence of such covariant formalism, there are four polarization degrees of freedom. They can be written in a four-dimensional tensor product space
\begin{eqnarray}\label{4a}
a_{1}=\textrm{a}_1\otimes 1\otimes 1\otimes 1,~~~~
a_{2}=1\otimes \textrm{a}_2\otimes 1\otimes 1,
\\
a_{3}=1\otimes 1\otimes \textrm{a}_3\otimes 1,~~~~
a_{0}=1\otimes 1\otimes 1\otimes \textrm{a}_0,
\end{eqnarray}
and such space is spanned by kets of the form
\begin{eqnarray}\label{ket4irr}
|n_1,n_2,n_3,n_0\rangle
&=&
|n_1\rangle \otimes |n_2\rangle \otimes |n_3\rangle \otimes |n_0\rangle.
\end{eqnarray}
Let us also take a closer look at
\begin{eqnarray}
a_{{\bf a}}^{\dagger} 
a^{{\bf a}}
&=&
\frac{1}{2}q_{{\bf a}}q^{{\bf a}}
+2
+\frac{1}{2}
p_{{\bf a}}p^{{\bf a}}.
\end{eqnarray}
This implies that the Hamiltonian operator (\ref{covH}) can be written in terms of operators $a_{{\bf a}}^{\dagger}$ and $a^{{\bf a}}$ in the form
\begin{eqnarray}
H=
-a_{{\bf a}}^{\dagger} 
a^{{\bf a}}
+2,
~~~~~ {\bf a}=0,1,2,3.
\end{eqnarray}
Furthermore, the following commutation relations hold
\begin{equation}
[H,a_{{\bf a}}]
=
[ -a_{{\bf {b}}}^{\dagger} 
a^{{\bf {b}}}
,a_{{\bf a}}]
=
-a_{{\bf a}},
\end{equation}
\begin{equation}
[H,a_{{\bf a}}^{\dagger}]
=
[- a_{{\bf {b}}}^{\dagger} 
a^{{\bf {b}}}
,a_{{\bf a}}^{\dagger}]
=
a_{{\bf a}}^{\dagger}.
\end{equation}
Let us also assume that the eigenvalue of the covariant Hamiltonian operator (\ref{covH}) acting on four-dimensional space is denoted by $E$. 
\begin{eqnarray}
H|n_1,n_2,n_3,n_0\rangle
&=&
E|n_1,n_2,n_3,n_0\rangle.
\end{eqnarray}
At this point this is really a quantity corresponding to the number of particles, but in the upcoming Section \ref{sec:4dreducible} an extension of this model to reducible representation algebra is made and $E(1)$ will correspond to the total energy operator.
For now it can be shown that indeed $a_{{\bf a}}$ lowers and $a_{{\bf a}}^{\dag}$ raises $E$ by 1, i.e.
\begin{eqnarray}\label{defanihilation}
Ha_{{\bf a}}|n_1,n_2,n_3,n_0\rangle
=
(E-1)a_{{\bf a}}|n_1,n_2,n_3,n_0\rangle,
\\
\label{defcreation}
Ha_{{\bf a}}^{\dag}|n_1,n_2,n_3,n_0\rangle\
=
(E+1)a_{{\bf a}}^{\dag}|n_1,n_2,n_3,n_0\rangle.
\end{eqnarray}
\subsection{Construction for space-like photons }\label{sec:4dconstruction123}
From the previous section we learn that (\ref{defanihilation}) and (\ref{defcreation}) define lowering and raising energy operators respectively and the raising operators will be denoted with a dagger. These operators are also know in the literature as annihilation and creation operators, but at this point this terminology will not be used, since the definition of the ${\bf a}=0$ polarization degree (time-like) may be ambiguous.

It is quite evident that for the polarizations indexed by ${\bf a}=1,2,3$, the raising energy operators create new states. Let us then define vacuum states for these three dimensions as normalized states that are annihilated by lowering energy operators
\begin{eqnarray}\label{vac123}
\textrm{a}_{{\bf{j}}}|0\rangle
&=&
0,~~~~{{\bf{j}}}=1,2,3.
\end{eqnarray}
Now let us normalize these space-like states to 1. The state of $n_{{\bf{j}}}$ excitations must be proportional to $n$ raising energy operators acting on ground state
\begin{eqnarray}
|n_{{\bf{j}}}\rangle
~\sim~
\textrm{a}_{{\bf{j}}}^{\dag n}|0\rangle
,~~~~{{\bf{j}}}=1,2,3.
\end{eqnarray}
Then the scalar product 
\begin{eqnarray}
\langle
n_{{\bf{i}}}|n_{{\bf{j}}}
\rangle
&~\sim~&
\nonumber\\
\langle 0|
\textrm{a}_{{\bf{i}}}^{n}
\textrm{a}_{{\bf{j}}}^{\dag n}|0\rangle
&=&
\delta_{{\bf{j}}}{_{{\bf{i}}}} n
\langle 0|
\textrm{a}_{{\bf{j}}}^{n-1}
\textrm{a}_{{\bf{i}}}^{\dag n-1}|0\rangle
\nonumber\\
&=&
\delta_{{\bf{j}}}{_{{\bf{i}}}} n(n-1)
\langle 0|
\textrm{a}_{{\bf{j}}}^{n-2}
\textrm{a}_{{\bf{i}}}^{\dag n-2}|0\rangle=...
\qquad
\end{eqnarray}
Going further with this recurrence, we get
\begin{eqnarray}
\langle
n_{{\bf{i}}}|n_{{\bf{j}}}
\rangle
~\sim~
\delta_{{\bf{i}}}{_{{\bf{j}}}} n!
\langle 0
|0\rangle.
\end{eqnarray}
Now we can give a normalized definition of bras and kets
\begin{eqnarray}
|n_{{\bf{j}}}\rangle
&=&
\frac{1}{\sqrt{n!}}
\textrm{a}_{{\bf{j}}}^{\dag n}|0\rangle,~~~~{\bf{j}}=1,2,3,
\nonumber\\
\langle n_{{\bf{j}}}|
&=&
\frac{1}{\sqrt{n!}}
\langle 0|
\textrm{a}_{{\bf{j}}}^n,~~~~{\bf{j}}=1,2,3.
\end{eqnarray}
This means that the action of raising and lowering operators is defined as follows
\begin{eqnarray}
\textrm{a}_{{\bf{j}}}^{\dagger}| n_{{\bf{j}}}\rangle  &=& \sqrt{n+1}| n_{{\bf{j}}}
+1\rangle,
\\
\textrm{a}_{{\bf{j}}}| n_{{\bf{j}}}\rangle &=& \sqrt{n}| n_{{\bf{j}}}-1\rangle.
\end{eqnarray}
We also define number operators for these three polarizations
\begin{eqnarray}~\label{n123}
\textrm{a}_{{\bf{j}}}^\dagger\textrm{a}_{{\bf{j}}}| n_{{\bf{j}}}\rangle 
&=& 
n_{{\bf{j}}}| n_{{\bf{j}}}\rangle,~~~~{\bf{j}}=1,2,3.
\end{eqnarray}
Furthermore, for the $q$ representation of vacuum (\ref{vac123}), we get a differential equation
\begin{eqnarray}\label{}
\frac{1}{\sqrt{2}}
\left(
q_{{\bf{j}}}+\frac{\partial}{\partial q_{{\bf{j}}}}
\right)
O(q_{{\bf{j}}})
=
0,~~~~{\bf{j}}=1,2,3,
\end{eqnarray}
with a simple solution
\begin{eqnarray}\label{}
O(q_{{\bf{j}}})
=
A\exp\left(
-\frac{q_{{\bf{j}}}^2}{2}
\right)
,~~~~{\bf{j}}=1,2,3.
\end{eqnarray}
Here $A$ is the normalization constant
\begin{eqnarray}\label{}
A
=
\left(
\frac{1}{\pi}
\right)^{\frac{1}{4}}.
\end{eqnarray}
The whole construction for these three polarization degrees of freedom is evident. Now without any confusion we can say that operators that raise the total energy, create particles, so can be called creation operators. Also those that lower the total energy should annihilate the vacuum. Not only the states have positive norms but also the vacuum in the canonical $q$ representation is proper, i.e. satisfies the requirement of going to zero at infinity, making it possible to normalize the function. 
\subsection{Construction for time-like photons, where lowering energy operators annihilate vacuum}\label{sec:4dconstruction0e+}
The problem arises for the ${\bf a}=0$ degree. The question for the time-like polarization is: do the lowering energy operators annihilate particles and raising energy operators create ones or maybe do the lowering energy operators create particles and raising energy operators annihilate them. 
First, let us assume that the lowering operator annihilates the vacuum, i.e.
\begin{eqnarray}\label{vac0}
\textrm{a}_{0}|0\rangle
&=&
0.
\end{eqnarray}
Then, the state of $n_{0}$ excitations is proportional to $n$ raising operators acting on the ground state, so that
\begin{eqnarray}
|n_{0}\rangle
~\sim~
\textrm{a}_{0}^{\dag n}|0\rangle
\end{eqnarray}
and the scalar product reads
\begin{eqnarray}
\langle
n_{0}|n_{0}
\rangle
&~\sim~&
\nonumber\\
\langle 0|
\textrm{a}_{0}^{n}
\textrm{a}_{0}^{\dag n}|0\rangle
&=&
n
\langle 0|(-)
\textrm{a}_{0}^{n-1}
\textrm{a}_{0}^{\dag n-1}|0\rangle
\nonumber\\
&=&
(-)^2 n(n-1)
\langle 0|
\textrm{a}_{0}^{n-2}
\textrm{a}_{0}^{\dag n-2}|0\rangle=...
\qquad
\end{eqnarray}
Going further with the recurrence we get
\begin{eqnarray}
\langle
n_{0}|n_{0}
\rangle
~\sim~
(-)^n n!
\langle 0
|0\rangle.
\end{eqnarray}
It looks like the states corresponding to odd values of $n$ give negative norms. Giving a normalized to 1 definition of bras and kets becomes a problem now. If we assume
\begin{eqnarray}
|n_{0}\rangle
&=&
\frac{1}{\sqrt{n!}}
(i \textrm{a}_{0}^{\dag })^n|0\rangle,
\nonumber\\
\langle n_{0}|
&=&
\frac{1}{\sqrt{n!}}
\langle 0|
\left(i \textrm{a}_{0}\right)^n,
\end{eqnarray}
then saving the positivity of the scalar product has an effect on the hermitian conjugate operation. 
On the other hand we could follow Gupta \cite{Gupta50} and leave the metric indefinite. 
But even then another problem arises, i.e. using the $q$ representation of vacuum, we get a differential equation
\begin{eqnarray}\label{}
\frac{1}{\sqrt{2}}
\left(
q_{0}-\frac{\partial}{\partial q_{0}}
\right)
O(q_{0})
=
0,
\end{eqnarray}
with a solution that is divergent at infinity, i.e.
\begin{eqnarray}\label{}
O(q_{0})
=
A\exp\left(
\frac{q_{0}^2}{2}
\right).
\end{eqnarray}
All these conclusions may suggest that another point of view on the time-like polarization is needed.
\subsection{Construction for time-like photons, where raising energy operators annihilate vacuum}\label{sec:4dconstruction0e-}
Now  let us assume that the raising operator annihilates the vacuum, which means that the energy spectrum is bounded from the top and to raise the total energy level we need to annihilate a particle, i.e.
\begin{eqnarray}\label{vac0}
\textrm{a}_{0}^{\dagger}|0\rangle
&=&
0.
\end{eqnarray}
Such a construction has lots of advantages which will be shown in this section.
Now the state of $n_{0}$ excitations is proportional to $n$ lowering energy operators acting on ground state, so that
\begin{eqnarray}
|n_{0}\rangle
~\sim~
\textrm{a}_{0}^{n}|0\rangle.
\end{eqnarray}
This means that creating new particles is lowering the total energy. Then the scalar product is
\begin{eqnarray}
\langle
n_{0}|n_{0}
\rangle
&~\sim~&
\nonumber\\
\langle 0|
\textrm{a}_{0}^{\dag n}\textrm{a}_{0}^{n}
|0\rangle
&=&
n
\langle 0|
\textrm{a}_{0}^{\dag n-1}
\textrm{a}_{0}^{n-1}|0\rangle
\nonumber\\
&=&
n(n-1)
\langle 0|
\textrm{a}_{0}^{\dag n-2}
\textrm{a}_{0}^{n-2}|0\rangle=...,
\end{eqnarray}
so that
\begin{eqnarray}
\langle
n_{0}|n_{0}
\rangle
~\sim~
n!
\langle 0
|0\rangle.
\end{eqnarray}
Giving a normalized definition of bras and kets is not a problem now
\begin{eqnarray}
|n_{0}\rangle
&=&
\frac{1}{\sqrt{n!}}
\textrm{a}_{0}^{n}|0\rangle,
\nonumber\\
\langle n_{0}|
&=&
\frac{1}{\sqrt{n!}}
\langle 0|
\textrm{a}_{0}^{\dag n}.
\end{eqnarray}
Furthermore, this means that for this representation the raising and lowering energy operators are defined as
\begin{eqnarray}
\textrm{a}_{0}| n_{0}\rangle  &=& \sqrt{n+1}| n_{0}+1\rangle,
\\
\textrm{a}_{0}^{\dagger}| n_{0}\rangle &=& \sqrt{n}| n_{0}-1\rangle.
\end{eqnarray}
It turns out that in this case we do not have to choose between the positivity of the scalar product and the hermitian conjugate operation. It should be stressed that the number operator for the time-like polarization should be defined carefully
\begin{eqnarray}~\label{n0}
\textrm{a}_{0}\textrm{a}_{0}^\dagger| n_{0}\rangle 
&=& 
n_{0}| n_{ 0}\rangle.
\end{eqnarray}
Now formulating the vacuum state is not problematic. Using the $q$ representation we get a differential equation
\begin{eqnarray}\label{}
\frac{1}{\sqrt{2}}
\left(
q_{0}+\frac{\partial}{\partial q_{0}}
\right)
O(q_{0})
=
0,
\end{eqnarray}
with a solution
\begin{eqnarray}\label{}
O(q_{0})
=
A\exp\left(
\frac{-q_{0}^2}{2}
\right).
\end{eqnarray}
Summarizing all the above, from now on we will use the $a_0^{\dag}$ for an operator that annihilates vacuum and $a_0$ for the creation operator of time-like polarization states.

Another interesting fact comes from such interpretation of the time-like photon. To see this we first split the Hamiltonian in two: for the transverse polarizations $H_{12}$ and the unmeasured ones $H_{03}$, such that 
\begin{eqnarray}\label{H1203}
H|n_1,n_2,n_3,n_0\rangle
=
\left(
H_{12}+H_{03}
\right)
|n_1,n_2,n_3,n_0\rangle,
\end{eqnarray}
where
\begin{eqnarray}
\label{H12}
&&\hspace{-6pt}H_{12}|n_1,n_2,n_3,n_0\rangle
\nonumber\\
&&=
-\frac{p_1p^1+q_1q^1+p_2p^2+q_2q^2}{2}
|n_1,n_2,n_3,n_0\rangle
\nonumber\\
&&=
\left(
n_{1}
+
n_{2}
+1
\right)
|n_1,n_2,n_3,n_0\rangle,
\\
\label{H03}
&&\hspace{-6pt}H_{03}|n_1,n_2,n_3,n_0\rangle
\nonumber\\
&&=
-\frac{p_0p^0+q_0q^0+p_3p^3+q_3q^3}{2}
|n_1,n_2,n_3,n_0\rangle
\nonumber\\
&&=
\left(
-n_{0}
+
n_{3}
\right)
|n_1,n_2,n_3,n_0\rangle.
\end{eqnarray}
From this we see that for such a definition of $a_0$, the energy of the ground state comes only from transverse polarizations.
\section{Four-dimensional oscillator reducible representation  algebra}\label{sec:4dreducible}
\subsection{Motivation for reducible representation algebra proposed by Czachor}\label{sec:redmot}
In 1925 Heisenberg, Born and Jordan postulated that energies of classical free fields look in Fourier space like oscillator ensembles. It should be stressed that at that time Heisenberg, Born and Jordan did not know the notation of Fock space and may not fully understood the role of eigenvalues of operators. Having to consider oscillators with different frequencies they considered one oscillator for each frequency mode. In such case the  ensemble had to be infinite, since the number of modes was infinite.

It is a well known problem that the standard canonical procedures for field quantization result in various infinity problems.
It was shown by Czachor \cite{MC00} that the assumption of having one oscillator for each frequency mode may not be natural. This assumption continued in a series of papers on reducible representation of CCR \cite{MC03} - \cite{MWMC0904}. The main idea for reducible representation algebra is that each oscillator is a wave packet, a superposition of infinitely many different momentum states. 
To describe this concept in more detail let us first introduce a spectral decomposition of the frequency operator $\Omega$  
\begin{eqnarray}\label{omega}
\Omega
&=&
\int
d \Gamma ({\bf k})~
\omega({\bf k})~
|{\bf k}\rangle\langle{\bf k}|,
\end{eqnarray}
so that (\ref{omega}) fulfills the following eigenvalue problem
\begin{eqnarray}
\Omega|{\bf k}\rangle
&=&
\int
d \Gamma ({\bf k}')~
\omega({\bf k}')
|{\bf k}'\rangle\langle{\bf k}'|
{\bf k}\rangle
=
\omega({\bf k})
|{\bf k}\rangle.
\end{eqnarray}
Here $d \Gamma ({\bf k})$ is the Lorentz invariant measure
\begin{equation}\label{Lormeasure}
d\Gamma({\bf{k}})
=\frac{d^3k}{(2\pi)^{3} 2 |{\bf k}|}
.
\end{equation}
Furthermore, kets of momentum are normalized to
\begin{equation}
\langle{\bf k}|{\bf k}'\rangle
=
(2\pi)^{3}2|{\bf k}|\delta^{(3)}
({\bf k},{\bf k}')
=
\delta_{\Gamma}({\bf k},{\bf k}'),
\end{equation} 
and the resolution of unity is
\begin{equation}
\int_{R^3}
d \Gamma ({\bf k})
|{\bf k}\rangle \langle {\bf k}|
=
I.
\end{equation}
The energy for photons, assuming the convention $\hbar=1$, is $E({\bf k})=\omega({\bf k})=|{\bf k}|$. Then the simplest Hamiltonian for one kind of polarization can be written in the form
\begin{eqnarray}
H
&=&
\Omega
\otimes
\left(
a^{\dag}a
+
\frac{1}{2}
\right)
\nonumber\\
&=&
\int
d \Gamma ({\bf k})~
\omega({\bf k})~
|{\bf k}\rangle\langle{\bf k}|
\otimes
\left(
a^{\dag}a
+
\frac{1}{2}
\right),
\end{eqnarray}
so that
\begin{eqnarray}
H
|{\bf k}, n
\rangle
&=&
\omega({\bf k})
\left(
n+\frac{1}{2}
\right)
|{\bf k}, n
\rangle.
\end{eqnarray}
Here ket $|n\rangle$ is the eigenvector of a ``standard theory" number operator 
$a^{\dagger}a$, where $[a,a^{\dagger}]=1$. 
Now let us introduce an operator that lives in both: the momentum and polarization spaces
\begin{eqnarray}\label{ared}
a({\bf k},1)
=
|{\bf k}\rangle\langle{\bf k}|
\otimes
a.
\end{eqnarray}
In bracket of $a({\bf k},1)$ in (\ref{ared}), ${\bf k}$ indicates that this representation is reducible and $1$ that it is the $N=1$ oscillator representation. Using the resolution of unity, we can also define an operator within the whole spectrum of frequencies
\begin{eqnarray}
a(1)
=
\int
d \Gamma ({\bf k})~
a({\bf k},1)
=
I
\otimes
a,
\end{eqnarray}
such that the commutator $[a(1),a^{\dagger}(1)]=I\otimes 1$. Here $1$ in the bracket of $a(1)$ denotes that this is the $N=1$ oscillator representation.
\subsection{Lie algerba}\label{sec:redladder}
The four-dimensional $N=1$ (or 1-oscillator) representation of CCR acts in the Hilbert space
${\cal H}(1)$ spanned by kets of the form
\begin{eqnarray}\label{ket41xy}
&&\hspace{-6pt}|{\bf k},n_1,n_2,n_3,n_0\rangle
\nonumber\\
&&\quad=
|{\bf k}\rangle\otimes
\frac{
	(a_1^{\dag})^{n_1}
	(a_2^{\dag})^{n_2}
	(a_3^{\dag})^{n_3}
	(a_0)^{n_0}}
{\sqrt{n_0!n_1!n_2!n_3!}}|0,0,0,0\rangle.
\quad
\end{eqnarray}
Operators $a_{1}, a_{2}, a_{3}, a_{0}$ satisfy the commutation relations typical for irreducible
representation of CCR (\ref{comaadag}). In (\ref{ket41xy}) $a_1^{\dag},~a_2^{\dag}$ stand for creation operators for linear polarized photons and $a_0$, $a_3^{\dag}$ are both creation operators for time-like and longitudinal photons respectively. 
For the reducible representation we define the ladder operators
\begin{equation}
a_{{\bf a}}({\bf k},1)
=
|{\bf k}\rangle\langle{\bf k}|\otimes a_{{\bf a}},
~~~~~ {\bf a}=0,1,2,3.
\end{equation}
Then the following CCR algebra holds
\begin{eqnarray}\label{CCR41}
[a_{{\bf a}}({\bf k},1),a_{{\bf {b}}}
({\bf k}',1)^{\dagger}]
&=&
-g_{{\bf {ab}}}
\delta_{\Gamma}({\bf k},{\bf k}')
|{\bf k}\rangle\langle{\bf k}|\otimes 1_4
\nonumber\\
&=&
-
g_{{\bf {ab}}}
\delta_{\Gamma}({\bf k},{\bf k}')
I({\bf k} ,1)
.
\end{eqnarray}
This algebra representation is reducible, since the right-hand side of the commutator (\ref{CCR41}) is an operator-valued distribution $I({\bf k},1)=|{\bf k}\rangle\langle{\bf k}|\otimes 1_4$ belonging to the center of algebra, i.e.
\begin{equation}
[a_{{\bf a}}({\bf k},1),I({\bf k}',1)]
=
[a_{{\bf a}}({\bf k},1)^{\dagger},I({\bf k}',1)]
=
0,
\end{equation}
where operator $I({\bf k},1)$ forms the resolution of unity for 
$\mathcal{H}(1)$ Hilbert space
\begin{equation}
\int d \Gamma({\bf k}) ~I({\bf k},1)
=I(1).
\end{equation}
The number operator for the reducible representation algebra in $\mathcal{H}(1)$ Hilbert space will be defined as
\begin{eqnarray}\label{n1}
n_{{\bf j}}({\bf k},1)
&=&
|{\bf k}\rangle\langle{\bf k}|\otimes a_{{\bf j}}^{\dagger}a_{{\bf j}},
~~~~~ {\bf j}=1,2,3,
\\
n_{0}({\bf k},1)
&=&
|{\bf k}\rangle\langle{\bf k}|\otimes a_{{0}}a_{{0}}^{\dagger}.
\end{eqnarray}
We can also define a number operator within the whole spectrum of frequencies as follows
\begin{eqnarray}\label{n(1)}
n_{{\bf a}}(1)
&=&
\int 
d\Gamma({\bf k})
n_{\bf a}({\bf k},1),
~~~~~ {\bf a}=0,1,2,3.
\end{eqnarray}
Then the eigenvalue definition of the number operator, i.e. the number of photons of one kind of polarization within the whole frequency spectrum, would be
\begin{eqnarray}
n_{{\bf a}}(1)
|{\bf k},n_1,n_2,n_3,n_0\rangle
=
n_{{\bf a}}
|{\bf k},n_1,n_2,n_3,n_0\rangle.
\end{eqnarray}
Now the following Lie algebra for the reducible representation holds
\begin{equation}
[a_{{\bf a}}({\bf k},1),
a_{{\bf b}}({\bf k}',1)^{\dagger}]
=
-g_{{\bf a}{\bf b}}
\delta_{\Gamma}({\bf k},{\bf k}')
I({\bf k},1),
\end{equation}
\begin{equation}
[a_{{\bf a}}({\bf k},1),
n_{{\bf b}}({\bf k}',1)]
=
-g_{{\bf a}{\bf b}}
\delta_{\Gamma}({\bf k}, {\bf k}')
a_{{\bf b}}({\bf k},1),
\end{equation}
\begin{equation}
[a_{{\bf a}}({\bf k},1)^{\dagger},
n_{{\bf b}}({\bf k}',1)]
=
g_{{\bf a}{\bf b}}
\delta_{\Gamma}({\bf k}, {\bf k}')
a_{{\bf b}}({\bf k},1)^{\dagger} .
\end{equation}
Furthermore, for the representation within the whole frequency spectrum we have
\begin{equation}
[a_{{\bf a}}(1),
a_{{\bf b}}(1)^{\dagger}]
~=
-
g_{{\bf a}{\bf b}}
I(1),
\end{equation}
\begin{equation}
[a_{{\bf a}}(1),
n_{{\bf b}}(1)]
~=
-
g_{{\bf a}{\bf b}}
a_{\bf b}(1),
\end{equation}
\begin{equation}
[a_{{\bf a}}(1)^{\dagger},
n_{{\bf b}}(1)]
~=
g_{{\bf a}{\bf b}}
a_{{\bf b}}(1)^{\dagger} .
\end{equation}
As we can see, the Lie algebra for the whole frequency spectrum has the ``standard theory" structure.
\subsection{Hamiltonian}\label{sec:redHam}
Now, let us introduce a covariant Hamiltonian in the reducible representation algebra
\begin{eqnarray}
H(1)\label{covredH}
&=&
\Omega
\otimes
H.
\end{eqnarray}
Here $\Omega$ is the frequency operator (\ref{omega}) and $H$ is defined in (\ref{covH}).
Hamiltonian (\ref{covredH}) can be also written in terms of operators $a_{{\bf a}}^{\dagger}$ and  $a_{{\bf a}}$ in the form
\begin{eqnarray}
H(1)
&=&
\int
d \Gamma({\bf k})
~|{\bf k}|~
|{\bf k}\rangle \langle{\bf k}|
\otimes
\left(
-a_{{\bf a}}^{\dagger} 
a^{{\bf a}}
+2
\right)
.
\end{eqnarray}
Now the following commutation relations hold for the reducible representation algebra
\begin{equation}
[H(1),a_{{\bf a}}({\bf k},1)]
=
-|{\bf k}|~a_{{\bf a}}({\bf k},1),
\end{equation}
\begin{equation}
[H(1),a_{{\bf a}}({\bf k},1)^{\dagger}]
=
|{\bf k}|~a_{{\bf a}}({\bf k},1)^{\dagger}.
\end{equation}
Moreover, within the whole frequency spectrum of the ladder operators, we get
\begin{equation}
[H(1),a_{{\bf a}}(1)]
=
-\int
d \Gamma({\bf k})~
|{\bf k}|~a_{{\bf a}}({\bf k},1)
=
-\Omega\otimes a_{{\bf a}},
\end{equation}
\begin{equation}
[H(1),a_{{\bf a}}(1)^{\dagger}]
=
\int
d \Gamma({\bf k})~
|{\bf k}|~a_{{\bf a}}({\bf k},1)^{\dagger}
=
\Omega\otimes a_{{\bf a}}^{\dag}.
\end{equation}
Let us also assume that the eigenvalue of the covariant Hamiltonian operator (\ref{covredH}) acting on the four-dimensional space of 1-oscillator reducible representation is the total energy denoted by $E(1)$
\begin{eqnarray}
&&\hspace{-6pt}H(1)|{\bf k}, n_1,n_2,n_3,n_0\rangle
\nonumber\\
&&=
E(1)|{\bf k}, n_1,n_2,n_3,n_0\rangle
\nonumber\\
&&=
|{\bf k}|~(n_1+n_2+n_3-n_0+1)|{\bf k}, n_1,n_2,n_3,n_0\rangle.\qquad
\end{eqnarray}
Now it can be shown that indeed $a_{{\bf a}}(1)$ lowers and $a_{{\bf a}}(1)^{\dag}$ raises the total energy by $|{\bf k}|$, i.e.
\begin{eqnarray}\label{defanihilationred}
&&\hspace{-6pt}H(1)a_{{\bf a}}(1)|{\bf k},n_1,n_2,n_3,n_0\rangle
\nonumber\\
&&=
(E(1)-|{\bf k}|)a_{{\bf a}}(1)|{\bf k},n_1,n_2,n_3,n_0\rangle,
\end{eqnarray}
\begin{eqnarray}
&&\hspace{-6pt}H(1)a_{{\bf a}}(1)^{\dag}|{\bf k},n_1,n_2,n_3,n_0\rangle\label{defcreationred}
\nonumber\\
&&=
(E(1)+|{\bf k}|)a_{{\bf a}}(1)^{\dag}|{\bf k},n_1,n_2,n_3,n_0\rangle.
\end{eqnarray}
\subsection{$N$-oscillator space}\label{sec:redladderN}
Now let us discuss an extension of this model to arbitrary $N$ oscillators. The parameter $N$ characterizes reducible representations but is not directly related to the number of photons. The Hilbert space for any $N$ oscillators reads
\begin{equation}
\mathcal{H}(N)
=
\underbrace{\mathcal{H}(1)\otimes\ldots\otimes\mathcal{H}(1)}_{N}
=
\mathcal{H}(1)^{\otimes N}.
\end{equation}
So that the $\mathcal{H}(N)$ Hilbert space is spanned by kets of the form
\begin{eqnarray}\label{ket41N}
|{\bf k}_1,n_1^{1},n_2^{1},n_3^{1},n_0^{1}
\rangle\otimes\dots\otimes |{\bf k}_N,n_1^{N},n_2^{N},n_3^{N},n_0^{N}\rangle.\qquad
\end{eqnarray}
Let us also define an operator
\begin{equation}\label{A^(n)}
A^{(n)}
=
\underbrace{I\otimes...\otimes I}_{n-1}
\otimes A\otimes\underbrace{I\otimes...\otimes I}_{N-n}.
\end{equation}
The top index $(n)$ shows the ``position" of the $A$ operator in $\mathcal H(N)$ space.
Then we construct the Hamiltonian for $N$ oscillators as follows
\begin{eqnarray}\label{covredHN}
H(N)
&=&
\sum_{n=1}^{N}
H(1)^{(n)}
.
\end{eqnarray}
A natural extension for the ladder operators to $N$-oscillator reducible representations would be
\begin{eqnarray}\label{aN}
a_{{\bf a}}({\bf k},N)
&=&
\frac{1}{\sqrt{N}}
\sum_{n=1}^{N}
a_{{\bf a}}({\bf k},1)^{(n)}.
\end{eqnarray}
Term $\frac{1}{\sqrt{N}}$ is the normalization factor for $N$-oscillator representations. The CCR algebras still hold
\begin{eqnarray}
[a_{{\bf b}}({\bf k},N),
a_{{\bf b}}({\bf k}',N)^{\dagger}]
=
-
g_{{\bf a}{\bf b}}
\delta_{\Gamma}({\bf k},{\bf k}')
I({\bf k},N),
\end{eqnarray}
where at the right-hand side there is an operator
\begin{equation}
I({\bf k},N)
=
\frac{1}{N}
\sum_{n=1}^N 
I({\bf k},1)^{(n)}
,
\end{equation}
which for all $N$ is also in the center of algebra, since
\begin{equation}
[a_{{\bf a}}({\bf k},N),I({\bf k}',N)]
=
[a_{{\bf a}}({\bf k},N)^{\dagger},I({\bf k}',N)]
=0.
\end{equation}
Then $I({\bf k},N)$ forms the resolution of unity for 
$\mathcal{H}(N)$ Hilbert space, i.e.
\begin{eqnarray}
\int d \Gamma({\bf k} )~I({\bf k},N)
=I(N)
=
\underbrace{I(1)\otimes...\otimes I(1)}_{N}
.\qquad
\end{eqnarray}
Again algebra representations within the whole frequency spectrum holds the ``standard theory" structure. 
Using definition (\ref{n1}) of the number operator for $N=1$, we write
\begin{eqnarray}\label{nIIIN}
n_{{\bf a}}({\bf k},N)
&=&
\sum_{n=1}^N n_{{\bf a}}({\bf k}, 1)^{(n)}.
\end{eqnarray}
Furthermore, the following Lie algebras for reducible $N$-oscillator representations hold
\begin{equation}
[a_{{\bf a}}({\bf k},N),
n_{{\bf b}}({\bf k}',N)]
=
-g_{{\bf a}{\bf b}}
\delta_{\Gamma}({\bf k}, {\bf k}')
a_{{\bf b}}({\bf k},N),
\end{equation}
\begin{equation}
[a_{{\bf a}}({\bf k},N)^{\dagger},
n_{{\bf b}}({\bf k}',N)]
=
g_{{\bf a}{\bf b}}
\delta_{\Gamma}({\bf k}, {\bf k}')
a_{{\bf b}}({\bf k},N)^{\dagger} .
\end{equation}
\subsection{Vacuum and energy of vacuum}\label{sec:4dvacuum}
The subspace of vacuum states is spanned by vectors of the  form
\begin{eqnarray}
|{\bf k}_1,0,0,0,0\rangle\otimes\dots\otimes |{\bf k}_N,0,0,0,0\rangle.
\end{eqnarray}
Vacuum in this representation is any state annihilated by any annihilation operator.
\begin{equation}\label{defvac}
a_{{\bf j}}(1)|O(1)\rangle=0,~~~~{\bf{j}}=1,2,3,
\end{equation}
\begin{equation}
a_{0}(1)^{\dagger}|O(1)\rangle=0.
\end{equation}
Therefore, in $N=1$ oscillator representation, we may write
\begin{equation}\label{vacN1}
|O(1)\rangle
=
\int d\Gamma({\bf k}) O({\bf k})
|{\bf k},0,0,0,0\rangle.
\end{equation}
From the normalization condition
\begin{equation}
\langle O(1)
|O(1)\rangle
=1
\end{equation}
we get 
\begin{equation}\label{sqrvac}
\int d\Gamma({\bf k}) |O({\bf k})|^{2} 
=
\int d\Gamma({\bf k}) Z({\bf k})
=
1.
\end{equation}
Here the scalar field $Z({\bf k})=|O({\bf k})|^{2} $ represents vacuum probability density. Furthermore, square integrability of (\ref{sqrvac}) implies that $Z({\bf k})$ must decay at infinity. This point is of special importance for reducible representation algebra quantization.  It turns out that regularization can be a consequence of employing such special form of scalar field in the definition of vacuum.

Recalling (\ref{H03}), where it is shown that the energy of the vacuum comes only from the two transverse polarization degrees of freedom, we calculate
\begin{eqnarray}\label{vacN1E}
H(1)|O(1)\rangle
&=&
\int d \Gamma({\bf k})
~|{\bf k}|~
O({\bf k})
|{\bf k},0,0,0,0\rangle.
\end{eqnarray}
The extension to $N$-oscillator space is assumed to be a tensor product of $N=1$ vacuum states, i.e.
\begin{equation}
|O(N)\rangle
=
|O(1)\rangle^{\otimes N}
=
\underbrace{|O(1)\rangle\otimes...\otimes|O(1)\rangle}_{N}.
\end{equation}
Of course, the normalization condition for the $N$-oscillator representations still holds
\begin{equation}
\langle O(N)
|O(N)\rangle
=
\langle O(1)
|O(1)\rangle^{N}
=1.
\end{equation}
Let us also take definition (\ref{covredHN}) for the Hamiltonian in $N$-oscillator representations and write
\begin{eqnarray}\label{vacNE}
H(N)|O(N)\rangle
&=&
\sum_{n=1}^N
H(1)
^{(n)}
|O(N)\rangle
.
\end{eqnarray}
We see that the expectation value of the vacuum energy  does not depend on the $N$ parameter, i.e.
\begin{eqnarray}\label{vacNE}
&&\hspace{-6pt}\langle O(N)|
H(N)|O(N)\rangle
\nonumber\\
&&=
\langle O(1)
|O(1)\rangle^{N-1}
\langle O(1)|
H(1)|O(1)\rangle
\nonumber\\
&&=
\int d \Gamma({\bf k})
~|{\bf k}|~
Z({\bf k}).
\end{eqnarray}
This means that the energy of vacuum is not zero and depends only on the vacuum probability density $Z({\bf k})$. For the vacuum energy to be convergent we must demand
\begin{eqnarray}\label{vaccond}
\int d \Gamma({\bf k})
~|{\bf k}|~
Z({\bf k})
<\infty
.
\end{eqnarray}
Therefore, the convergence of vacuum is guaranteed by the proper choice of the vacuum probability density function and does not require the $N$ parameter at all. Furthermore, this analysis shows that the $N$ parameter may even be a finite number.
\subsection{Vector space corresponding to Maxwell's theory}\label{sec:4dHem}
Let us start by presenting the potential operator in reducible representation algebra for $N=1$ oscillator
\begin{eqnarray}\label{potN1}
A_{a}(x,1)
&=&
i\int d\Gamma({\bf k})
g_{a}{^{\bf a}}({\bf k})a_{{\bf a}}({\bf k},1) 
e^{-ik\cdot x}+\rm{H.c.}\qquad
\end{eqnarray}
Here, not just two operators corresponding to the polarizations, but four types are introduced. The same four-dimensional quantization, where in the place of the annihilation operator of the Gupta-Bleuler type potential for the time-like degree stands a creation operator, was already formulated by Czachor and Naudts \cite{MCJN05} and further by Czachor and Wrzask \cite{MCKW09}. Let us point out that $a_0$ is indeed a lowering total energy operator which is a creation operator like in \cite{MCKW09}, only it is denoted here, on the contrary to mentioned papers, without a dagger. Notation in this paper comes straightforward from the analysis in Section \ref{sec:4dconstruction0e-} and is convenient for the possibility of writing collective formulas. In (\ref{potN1}) $g_{a}{^{\bf a}}(\bf k)$ is a field of Minkowski tetrads. More on the Minkowski and null tetrad in Penrose notation is presented in \ref{sec:AppA}. Furthermore, in \ref{sec:AppE} we show that the potential operator (\ref{potN1}) transforms as a four-vector.

The four-divergence of the potential operator in $N=1$ oscillator representation reads
\begin{eqnarray}
&&\hspace{-6pt}\partial^{a}A_a(x,1)\label{lorenzA}
\nonumber\\
&&=
\frac{1}{\sqrt{2}}
\int d\Gamma({\bf k}) 
\left( 
a_0({\bf k},1) 
-
a_3({\bf k},1)
\right) 
e^{-ik\cdot x}+{\rm H.c.}\qquad\quad
\end{eqnarray}
As we can see this does not correspond to classical Maxwell theory, that is the Lorenz condition does not hold on the four-vector potential operator. Operator $a_0({\bf k},1)-a_3({\bf k},1)$ in D\"urr and Rudolph paper \cite{HPDER69} is called a ``bad ghost". This is a very adequate name because it spoils the classical electrodynamics correspondence. 

Returning to our problem, it would be appreciable for the theory to eliminate ``bad ghosts" and this can be done by solving a weaker, i.e. in averages Lorenz condition
\begin{eqnarray}\label{weeklorenz}
\langle \Psi_{EM}(1)|
\partial^{a}A_a(x,1)
|\Psi_{EM}(1) \rangle
&=&0.
\end{eqnarray}
We will try to impose this condition on the Hilbert space instead of on the operators. Here $\Psi_{EM}(1)$ are vectors that satisfy (\ref{weeklorenz}). Later, it will be shown that such vectors give equivalence to standard free-field Maxwell electromagnetism theory. From (\ref{weeklorenz}) it follows that
\begin{eqnarray}\label{weeklorenz2}
\langle\Psi_{EM}(1)|
\int d\Gamma({\bf k}) 
\left(  
a_0({\bf k},1) 
-
a_3({\bf k},1)
\right)
|\Psi_{EM}(1) \rangle
=0.
\nonumber\\
\end{eqnarray}
Solving (\ref{weeklorenz2}) we find that $\Psi_{EM}(1)$  has a coherent-like structure. 
Therefore, we define a displacement-like operator for time-like and longitudinal photons. We will start from the $N=1$ oscillator representation denoting
\begin{eqnarray}\label{DEMN1}
&&\hspace{-6pt}{\cal D}_{03}(\alpha,1)
\nonumber\\
&&=
\exp
\left(
\int d\Gamma({\bf k})
\Big(
\overline{\alpha({\bf k})}
(a_{0}({\bf k},1)
+
a_{3}({\bf k},1))
-
{\rm H.c.}
\Big)
\right).
\nonumber\\
\end{eqnarray}
Here $\alpha({\bf k})$ is a function corresponding to the ``amount of displacement" and can in general be dependent on ${\bf k}$.
Acting with operator (\ref{DEMN1}) on vacuum we get a vector state of the form
\begin{eqnarray}
&&\hspace{-6pt}
|\Psi_{EM}(1) \rangle=
{\cal D}_{03}(\alpha,1)|O(1)\rangle
\nonumber\\
&&=
\sum_{n_0,n_3=0}^{\infty}
\int d\Gamma({\bf k})
O({\bf k})
\exp
\left(-
|\alpha({\bf k})|^2
\right)
\nonumber\\
&&\hspace{8pt}\times
\frac{
	(\alpha({\bf k}))^{n_3} (\overline{\alpha({\bf k})})^{n_0}}
{\sqrt{n_3!n_0!}}
|{\bf k},n_1,n_2,n_3,n_0\rangle.
\end{eqnarray}
Operator (\ref{DEMN1}) shifts the ladder operators by $\alpha({\bf k})$
\begin{eqnarray}
&&\hspace{-6pt}
{\cal D}_{03}(\alpha,1)^{\dag}a_0({\bf k},1)
{\cal D}_{03}(\alpha,1)
=
a_0({\bf k},1)
+
\alpha({\bf k})I({\bf k},1),
\nonumber\\
\\
&&\hspace{-6pt}
{\cal D}_{03}(\alpha,1)^{\dag}a_3({\bf k},1)
{\cal D}_{03}(\alpha,1)
=
a_3({\bf k},1)
+
\alpha({\bf k})I({\bf k},1).
\nonumber\\
\end{eqnarray}
Therefore,
\begin{eqnarray}
&&\hspace{-6pt}
{\cal D}_{03}(\alpha,1)^{\dag}
\left(
a_0({\bf k},1)
-
a_3({\bf k},1)
\right)
{\cal D}_{03}(\alpha,1)
\nonumber\\
&&=
a_0({\bf k},1)
-
a_3({\bf k},1),
\end{eqnarray}
so that the week Lorenz condition (\ref{weeklorenz2}) holds. Furthermore, for $\Psi_{EM}(1)$ the number of time-like photons is equal to the number of longitudinal ones
\begin{eqnarray}\label{eqnon31}
\langle\Psi_{EM}(1)|
\left(  
n_0({\bf k},1) 
-
n_3({\bf k},1)
\right)
|\Psi_{EM}(1) \rangle
&=&0.\qquad
\end{eqnarray}
Therefore, their contribution cancels against each other.

The extension to arbitrary $N$-oscillator representations can be made by
\begin{eqnarray}
|\Psi_{EM}(N)\rangle
&=&
\underbrace{|\Psi_{EM}(1)\rangle\otimes...\otimes|\Psi_{EM}(1)\rangle}_{N}
=
|\Psi_{EM}(1)\rangle^{\otimes N}
,\nonumber\\
\end{eqnarray}
and the week Lorenz condition holds also for $N$-oscillator representations
\begin{eqnarray}\label{weeklorenzN}
&&\hspace{-6pt}
\langle\Psi_{EM}(N)|
\int d\Gamma({\bf k})
\left(
a_0({\bf k},N) 
-
a_3({\bf k},N)
\right)
|\Psi_{EM}(N) \rangle
\nonumber\\
&&\hspace{4pt}=0.
\end{eqnarray}
Also for arbitrary $N$ oscillators, the number of longitudinal photons equals to the number of time-like ones, i.e.
\begin{eqnarray}\label{n_0n_3N}
\langle\Psi_{EM}(N)|
\left(
n_0({\bf k},N) 
-
n_3({\bf k},N)
\right)
|\Psi_{EM}(N) \rangle
&=&0.\qquad
\end{eqnarray}

Let us note that this is not the usual Gupta-Bleuler condition such that
\begin{eqnarray}\label{guptacon2}
\left(
a_0
-
a_3
\right)
|\Psi \rangle
&=&0,
\end{eqnarray}
due to two aspects: a different definition of $a_0$ ladder operator and because it does not hold on the vector states but on the inner product. 
\subsection{Electromagnetic field operator}\label{sec:4dF}
The electromagnetic field operator for $N=1$ oscillator representation is by definition 
\begin{eqnarray}\label{e-mdef}
F_{ab}(x,1)
=
\partial_{a}A_{b}(x,1)-\partial_{b}A_{a}(x,1).
\end{eqnarray}
This can be written explicitly as
\begin{eqnarray}\label{e-m1}
&&\hspace{-6pt}
F_{ab}(x,1)
\nonumber\\
&&=
2\int d\Gamma({\bf k})
k_{[a}({\bf k})
g_{b]}{^{\bf{a}}}({\bf k})
a_{\bf{a}}({\bf k},1)
e^{-ik\cdot x}
+{\rm H.c.}
\nonumber\\
&&=\label{Fabtwofields}
\int d\Gamma({\bf k})
\left[
-k_a({\bf k})
x_{b}({\bf k}) 
+
k_{b}({\bf k})
x_{a}({\bf k})
\right]
a_{1}({\bf k},1) 
e^{-ik\cdot x}
\nonumber\\
&&+
\int d\Gamma({\bf k})
\left[
-k_a({\bf k})
y_{b}({\bf k}) 
+
k_{b}({\bf k})
y_{a}({\bf k})
\right]
a_{2}({\bf k},1) 
e^{-ik\cdot x}
\nonumber\\
&&+
\frac{1}{\sqrt{2}}
\int d\Gamma({\bf k})
\left[
t_a({\bf k})
z_{b}({\bf k}) 
-
t_{b}({\bf k})
z_{a}({\bf k})
\right]
\nonumber\\
&&\qquad\times
\left(
a_{0}({\bf k},1) 
-
a_{3}({\bf k},1) 
\right)
e^{-ik\cdot x}
+{\rm H.c.}
\end{eqnarray}
Here $k_a({\bf k})$ is the element of the null tetrad introduced in \ref{sec:AppA}. 
The field tensor (\ref{Fabtwofields}) can be split into two parts: consisting the transverse photon operators and the ``bad ghost" operators corresponding to particles unmeasured in experiments. It can be shown that in $\Psi_{EM}(1)$ the electromagnetic field operator corresponds to standard electromagnetism theory.
Furthermore, let us also check the free Maxwell equations for the electromagnetic field operator
\begin{eqnarray}
&&\hspace{-6pt}
\partial_c F_{ab}(x,1)\label{maxwelleq}
+
\partial_a F_{bc}(x,1)
+
\partial_b F_{ca}(x,1)
=
0,
\\
&&\hspace{-6pt}
\partial_{a}\label{maxwelleq2}
F^{ab}(x,1)
\nonumber\\
&&
=
\frac{i}{\sqrt{2}}
\int d\Gamma({\bf k})
k^{b}({\bf k})
\left(
a_{0}({\bf k},1) 
- 
a_{3}({\bf k},1) 
\right)
e^{-ik\cdot x}+{\rm H.c.}
\nonumber\\
\end{eqnarray}
The second equation (\ref{maxwelleq2}) consists of a ``bad ghost", but in $\Psi_{EM}(1)$ averages 
\begin{eqnarray}
\langle \Psi_{EM}(1)|
\partial_{a}
F^{ab}(x,1)
|\Psi_{EM}(1) \rangle
&=&
0.
\end{eqnarray}
As in the potential operator, the extension of the electromagnetic field to $N$-oscillator representations is equivalent to the extension for $N$ representations of the creation and annihilation operators in (\ref{e-m1}).
~~\\
\section{Lorentz transformation}\label{sec:Lorentz}
\subsection{Bogoliubov type transformation}\label{sec:LorentzBogoliubov}
In \ref{sec:AppB} and \ref{sec:AppC} transformation properties of spin-frames and the field of Minkowski tetrads are introduced in detail. In this section we will concentrate on the corresponding transformation on the ladder operators. 
At this point only the irreducible representation of CCR will be considered. Let us first define new ladder operators
\begin{equation}
b_{\bf{a}}
=
L_{\bf{a}}{^{\bf b}}(\Theta,\phi)a_{\bf{b}}.
\end{equation}
Transformation matrix $L_{\bf{a}}{^{\bf {b}}}(\Theta,\phi)$ can be written explicitly in terms of the Wigner phase $\Theta(\Lambda,{\bf k})$ and parameter  $\phi({\bf k})=|\phi({\bf k})|e^{i\xi({\bf k})}$ related to the gauge transformation. This is shown in (\ref{Labap}).
$L_{\bf{a}}{^{\bf b}}(\Theta,\phi)$ has the property of leaving the metric invariant, i.e.
\begin{equation}
g_{\bf{ab}}
=
L_{{\bf{a}}}{^{\bf{c}}}
(\Theta,\phi)
L_{\bf{b}}{^{\bf{d}}}
(\Theta,\phi)
g_{\bf{cd}},
\end{equation}
so that the new operators satisfy the same CCR
\begin{eqnarray}
[b_{\bf{a}},b_{\bf{b}}^{\dagger}]
&=&
[
L_{\bf{a}}{^{\bf {c}}}(\Theta,\phi)a_{\bf{c}}
~,
L_{\bf{b}}{^{\bf {d}}}(\Theta,\phi)a_{\bf{d}}^{\dagger}
]
\nonumber\\
&=&
-L_{\bf{a}}{^{\bf {c}}}(\Theta,\phi)
L_{\bf{b}}{^{\bf {d}}}(\Theta,\phi)
g_{\bf{cd}}
=
-g_{\bf{ab}}.\qquad
\end{eqnarray}
Therefore, there must exist an unitary Bogoliubov-type transformation $U(\Theta,\phi) $ satisfying
\begin{eqnarray}\label{bogulubovtrans}
b_{\bf{a}}
=
U(\Theta,\phi)^{\dag}
a_{\bf{a}}
U(\Theta,\phi)
=
L_{\bf{a}}{^{\bf b}}(\Theta,\phi)
a_{\bf{b}}.
\end{eqnarray}
In the next section an explicit representation of  $U(\Theta,\phi)$ will be given.
\subsection{Wigner rotations and gauge transformation in four-dimensional polarization space}\label{sec:Lorentzrepresentation}
In 1939 Wigner studied the subgroups of the Lorentz group, whose transformations leave the four-momentum of a given free particle invariant \cite{EPW39}. The maximal subgroup of the Lorentz group, which leaves the four momentum invariant is called the little group. This implies that the little group governs the internal space-time symmetries of relativistic particles. Wigner showed in his paper that invariant space-time symmetries are dictated by O(3)-like little groups in the case of massive particles and by E(2)-like little groups in the case of the massless ones. The application for photons has been discussed in many papers, among all \cite{JJ71}--\cite{HKS86}. It is also known that the Lorentz group is a very natural language for polarized light.
In order to explicitly construct transformation $U(\Theta,\phi)$ in (\ref{bogulubovtrans}), let us first introduce the representation of the Lie algebra generators of rotations $J_{\bf i}$ around an ${\bf i}$-th axis. This first will be done in canonical variables (\ref{canvar}) introduced earlier
\begin{equation}
J_{\bf i}=\varepsilon_{\bf {ijk}}q_{\bf j}p_{{\bf k}}.
\end{equation}
Using formulas (\ref{acov})-(\ref{acovdag}) we can write the generators of rotations in terms of the ladder operators 
\begin{eqnarray}
&J_{\bf i}&=-i\varepsilon_{\bf {ijk}}a_{\bf j}^{\dagger}a_{{\bf k}}.
\end{eqnarray}
Also let us introduce boosts $K_{\bf i}$ along an ${\bf i}$-th axis, first in terms of canonical variables (\ref{canvar})
\begin{equation}
K_{\bf i}=p_{\bf i}q_{0}-p_{0}q_{\bf i}.
\end{equation}
Again using formulas (\ref{acov})-(\ref{acovdag}) we write the generators of boosts in terms of the ladder operators of general form
\begin{eqnarray}
&K_{\bf i}&=i (a_{\bf i}^{\dagger}a_{0}
-a_{0}^{\dagger}a_{\bf i}).
\end{eqnarray}
Interestingly, the same form of generators was introduced earlier in \cite{MCJN05}, although  without the knowledge of canonical variables (\ref{canvar}). These generators satisfy the following commutation relations
\begin{equation}
[J_{\bf i},J_{\bf j}]
=
i\varepsilon_{\bf {ijk}}J_{{\bf k}},
\end{equation}
\begin{equation}
[K_{\bf i},K_{\bf j}]
=
-i\varepsilon_{\bf {ijk}}J_{{\bf k}},
\end{equation}
\begin{equation}
[J_{\bf i},K_{\bf j}]
=
i\varepsilon_{\bf{ijk}}K_{{\bf k}}.
\end{equation}
Wigner in his paper \cite{EPW39} showed that the little group for massless particles moving along $z$ axis is generated by the rotation generators around $z$ axis $J_3$ and two other generators. These other two generators are combinations of $J_{\bf i}$ and $K_{\bf i}$ and form a representation of an Euclidean group E(2), i.e.
\begin{equation}\label{L1L2}
L_{1}=J_{1}+K_{2},
\qquad
L_{2}=J_{2}-K_{1},
\qquad
L_{3}=J_{3},
\end{equation}
with the following commutation relations
\begin{equation}
[L_{1},L_{3}]=-iL_{2}, \quad
[L_{2},L_{3}]=iL_{1}, \quad
[L_{1},L_{2}]=0.
\end{equation}
The physical variable associated with $J_3$ is the helicity degree of freedom of massless particles, but it was not clear what is the physical interpretation of generators $L_1$ and $L_2$. In 1971 Janner and Janssen \cite{JJ71} showed that those generators generate translations and are responsible for gauge transformations of the four potential. This will be discussed further in Section \ref{sec:gaugetransfor}.
Moreover, it should be stressed that only $L_3$ annihilates the vacuum states, since it is normally ordered, contrary to generators $L_1$ and $L_2$.

Now the Bogoliubov-type transformation can be written as
\begin{equation}\label{U(irr)}
U(\Theta,\phi)
=
\exp(i\alpha_{1}L_{1}+i\alpha_{2}L_{2}+i\alpha_{3}L_{3})
,
\end{equation}
with parameters
\begin{equation}
\alpha_{1}(\Theta,\phi)
=
-
\frac{\Theta(\Lambda, {\bf k})}{\sin\Theta(\Lambda, {\bf k})}
|\phi({\bf k})|\sin (\xi({\bf k})+\Theta(\Lambda, {\bf k})),
\end{equation}
\begin{equation}
\alpha_{2}(\Theta,\phi)
=
-
\frac{\Theta(\Lambda, {\bf k})}{\sin\Theta(\Lambda, {\bf k})}
|\phi({\bf k})|\cos(\xi({\bf k})+\Theta(\Lambda, {\bf k})),
\end{equation}
\begin{equation}
\alpha_{3}(\Theta)
=
-
2\Theta(\Lambda, {\bf k}).
\end{equation}
As we can see the parameter $\alpha_3$ depends only on the Wigner phase. 
We can also write formula (\ref{U(irr)}) in following  forms
\begin{eqnarray}
&&\hspace{-6pt}U(\Theta,\phi)
\nonumber\\
&&=
\exp{\left(-i|\phi| \sin(\xi+2\Theta) L_1-i|\phi| \cos(\xi+2\Theta) L_2\right)}
\nonumber\\
&&\quad\times	
\exp{\left(-2 i \Theta L_3\right)}\label{UGUR}
\\
&&=
\exp{\left(-2 i\Theta L_3\right)}\label{URUG}
\nonumber\\
&&\quad\times
\exp{\left(-i |\phi| \sin \xi L_1
	-i |\phi| \cos \xi L_2\right)}
\end{eqnarray}
and define transformations associated with the Lorentz transformation and gauge transformation respectively
\begin{equation}	
U(\Lambda,{\bf k})
=
\exp{\left(-2 i\Theta(\Lambda, {\bf k}) L_3\right)},\label{UR}
\end{equation}	
\begin{eqnarray}	
U_G({\bf k})\label{UG}
=
\exp{\left(-i |\phi({\bf k})| \sin \xi({\bf k}) L_1
	-i |\phi({\bf k})| \cos \xi({\bf k}) L_2\right)},
\hspace{-1cm}
\nonumber\\
\end{eqnarray}	
such that
\begin{equation}
U(\Lambda,{\bf k})^{\dag}
a_{\bf{a}}
U(\Lambda,{\bf k})
=
R_{{\bf{a}}}{^{\bf{b}}}(\Lambda,{\bf k})
a_{\bf{b}},
\end{equation}	
\begin{equation}
U_G({\bf k})^{\dag}
a_{\bf{a}}
U_G({\bf k})
=
G_{{\bf{a}}}{^{\bf{b}}}({\bf k})
a_{\bf{b}},
\end{equation}
where $R_{{\bf{a}}}{^{\bf{b}}}(\Lambda,{\bf k})$ and 
$G_{{\bf{a}}}{^{\bf{b}}}({\bf k})$ are transformation matrices (\ref{Rab}) and (\ref{Gab}) introduced in \ref{sec:AppC}. 
\subsection{Lorentz transformation for the reducible representation algebra}\label{sec:Lorentzreducible}
To construct a Lorentz transformation for the reducible $N=1$ oscillator
representation the Bogoliubov-type transformation must be written as
\begin{equation}	
U(\Lambda,0,N=1)\label{LorentzTransN1}\label{WU}
=
\int d\Gamma({\bf k})
|{\bf k}
\rangle\langle
{\bf{\Lambda^{-1}k}}|
\otimes
U(\Lambda,{\bf k}).
\end{equation}	
Then, for the hermitian conjugate we can write
\begin{equation}\label{WU}
U(\Lambda,0,N=1)^{\dagger}
=
\int d\Gamma({\bf k})
|{\bf{\Lambda^{-1}k}}
\rangle\langle
{\bf k}|
\otimes U(\Lambda,{\bf k})^{\dag}.
\end{equation}	
Let us denote
\begin{eqnarray}\label{Wop}
W(\Lambda)
=
\int d\Gamma({\bf k})
|{\bf k}
\rangle\langle
{\bf{\Lambda^{-1}k}}|.
\end{eqnarray}
This operator is not dependent on spin and acts only on momentum ${\bf k}$, i.e.
\begin{eqnarray}
W(\Lambda)|{\bf {k}}\rangle
=
|{\bf{\Lambda k}}\rangle.
\end{eqnarray}
Operator (\ref{Wop}) leaves the inner product invariant and therefore is an unitary operator. The hermitian conjugate of (\ref{Wop}) is
\begin{eqnarray}
&&\hspace{-6pt}W(\Lambda)^{\dag}\label{Wopdag}
=
W(\Lambda^{-1})
=
\int d\Gamma({\bf k})
|{\bf{\Lambda^{-1}k}}
\rangle\langle
{\bf k}|,
\\
&&\hspace{-6pt}	W(\Lambda)^{\dag}|{\bf {k}}\rangle
=
|{\bf{\Lambda^{-1}k}}\rangle.
\end{eqnarray}
For the $N$-oscillator extension we can write
\begin{eqnarray}
U(\Lambda,0,N)\label{LorentzTransN}
=
U(\Lambda,0,1)^{\otimes N}.
\end{eqnarray}
Then, the transformation rule for the ladder operators in $N$-oscillator reducible representations
\begin{eqnarray}\label{translora}
&&\hspace{-6pt}U(\Lambda,0,N)^{\dagger}
a_{\bf{a}}({\bf k},N)
U(\Lambda,0,N)
\nonumber\\
&&=
R_{{\bf{a}}}{^{\bf{b}}}(\Lambda,{\bf k})
a_{\bf{b}}({\bf{\Lambda^{-1}k}},N).
\end{eqnarray}
Furthermore, it is shown in \ref{sec:AppD} that 
\begin{eqnarray}\label{Ulambdalambda'}
&&	U(\Lambda,0,1)U(\Lambda',0,1)
=
U(\Lambda\Lambda',0,1).
\end{eqnarray}
\subsection{Transformation properties of vacuum}\label{sec:Lorentzvacuum}
Let us remind ourselves that the definition of vacuum in this representation is not unique, i.e.
\begin{eqnarray}
|O(1)\rangle
=
\int d\Gamma({\bf k}) O({\bf k})
|{\bf k},0,0,0,0\rangle.
\nonumber
\end{eqnarray}
Then the Lorentz transformation acts on vacuum as follows
\begin{eqnarray}\label{vacuumtrans}
U(\Lambda,0,1)|O(1)\rangle
&=&
\int d\Gamma({\bf k}) O({\bf{\Lambda^{-1}k}})
|{\bf k},0,0,0,0\rangle
.\qquad
\end{eqnarray}
The transformed vacuum state is again a vacuum state, but the probability of finding ${\bf k}$ is modified by the \linebreak Doppler effect. As a byproduct we observe that the vacuum field transforms as a
scalar field
\begin{eqnarray}
O({\bf k})
~~\mapsto~~
O({\bf{\Lambda^{-1}k}})
.
\end{eqnarray}
This also implies the following transformation rule of the vacuum probability density
\begin{eqnarray}
Z({\bf k})
~~\mapsto~~
Z({\bf{\Lambda^{-1}k}})
.
\end{eqnarray}
Of course, the norm of such transformed vacuum, due to the Lorentz invariant measure (\ref{Lormeasure}), is invariant
\begin{eqnarray}
&&\hspace{-6pt}\langle O(1)| U(\Lambda,0,1)^{\dagger} U(\Lambda,0,1)
|O(1)\rangle 
\nonumber\\
&&=
\int d\Gamma({\bf k}) |O({\bf{\Lambda^{-1}k}})|^{2}  =1.
\qquad
\end{eqnarray}
\subsection{Gauge transformation}\label{sec:gaugetransfor}
Quantum field theory is assumed to be gauge invariant. The change of gauge is a change in electromagnetic potential that does not produce any change in physical observables. We will show that for the covariant reducible representation algebra there exists a transformation that corresponds to a gauge transformation of the potential in $\Psi_{EM}(1)$ vector space.  
Now let us start from the potential operator (\ref{potN1}) and see, how it transforms after transformation 
\begin{equation}
U_G(\phi)=
\int
d \Gamma ({\bf k})
|{\bf k}\rangle \langle {\bf k}|
\otimes
U_G({\bf k})^{\dag} .
\end{equation}
Then
\begin{eqnarray}\label{potN1gauge3}
\tilde{A}_{a}(x,1)
&=&
U_G(\phi)^{\dag} 
A_{a}(x,1)
U_G(\phi)
\nonumber\\
&=&
A_{a}(x,1)
+
\partial_a \varphi(x,1)
+
B_a (x,1).
\end{eqnarray}
Here
\begin{eqnarray}\label{potN1gauge4}
&&\hspace{-6pt}\varphi(x,1)
=
\sqrt{2}
\int d\Gamma({\bf k})
|\phi({\bf k})|
\left(
\cos\xi a_1({\bf k},1)\right.
\nonumber\\
&&\hspace{36pt}
\left.
+
\sin\xi a_2({\bf k},1)
\right) 
e^{-ik\cdot x} 
+\rm{H.c.},
\\
&&\hspace{-6pt}
B_a (x,1)\label{potN1gauge5}
=
i\int d\Gamma({\bf k})
\Big(
-x_{a}({\bf k})
|\phi|\cos \xi
+
y_{a}({\bf k}) 
|\phi|\sin \xi
\nonumber\\
&&\hspace{40pt}
+
\left( 
z_{a}({\bf k}) 
-
t_{a}({\bf k}) 
\right)
\frac{|\phi|^2}{2}
\Big)
\nonumber\\
&&\hspace{40pt}
\times
\left(a_3({\bf k},1) -a_0({\bf k},1)
\right)
e^{-ik\cdot x} +\rm{H.c.}
\hspace*{-20pt}
\end{eqnarray}
The $B_a(x,1)$ term contains the ``bad ghost" operator, but in $\Psi_{EM}(1)$ vector space this contribution vanishes, i.e.
\begin{eqnarray}\label{}
&&\hspace{-6pt}\langle \Psi_{EM}(1)|
\tilde{A}_{a}(x,1)
|\Psi_{EM}(1) \rangle
\nonumber\\
&&=
\langle \Psi_{EM}(1)|
\left(
A_{a}(x,1)
+
\partial_a \varphi(x,1)
\right)
|\Psi_{EM}(1) \rangle.
\end{eqnarray}
Moreover, the potential operator (\ref{potN1gauge3}) under transformation (\ref{UG}) holds the weaker Lorenz condition, i.e.
\begin{eqnarray}\label{potN1Lorenzgauge}
\partial^a
\tilde{A}_{a}(x,1)
=
\partial^a
A_{a}(x,1),
\end{eqnarray}
so that
\begin{eqnarray}
\langle \Psi_{EM}(1)|
\partial^{a}\tilde{A}_{a}(x,1)
|\Psi_{EM}(1) \rangle
&=&0.
\end{eqnarray}
Earlier, in Section \ref{sec:4dHem}, it was shown that there exists a space denoted by $\Psi_{EM}(1)$ in which a weaker Lorenz condition (\ref{weeklorenz}) holds. This result should be compared with those of Janner and Janssen \cite{JJ71} followed by Han, Kim and Son \cite{HKS82}. They worked out a similar conclusion that $L_1$ and $L_2$ generators carry gauge transformations for the potential operator $(A_0, A_1, A_2, A_3)$, where $A_0=A_3$. Here the conclusion is the same for equal numbers of longitudinal and time-like photons, i.e. with $n_0=n_3$. 
\subsection{Invariants in a combined homogeneous Lorentz and gauge transformation}\label{secmixing}
The combined homogeneous Lorentz and gauge transformation $U(\Theta,\phi)$ mixes transverse ladder operators $a_1,~a_2$ with excitations $a_3, ~a_0$. 
Let us first take a closer look at the ``bad ghost" operator $a_{3}-a_0$, i.e. 
\begin{eqnarray}
U(\Theta,\phi)^{\dag}  (a_{3}-a_0) U(\Theta,\phi)
&=&
a_{3} -a_{0}.
\end{eqnarray}
As we can see, this operator is invariant under transformation (\ref{U(irr)}). 
It is also easy to show that the covariant total number of photons does not change due to the combined Lorentz and gauge transformation, i.e.
\begin{eqnarray}
&&\hspace{-6pt}U(\Theta,\phi)^{\dag}
\left(
n_1+n_2+n_3-n_0
\right)
U(\Theta,\phi)
\nonumber\\
&&=
n_1+n_2+n_3-n_0.
\end{eqnarray}
Furthermore for the Lorentz transformation we get
\begin{eqnarray}
&&\hspace{-6pt}U(\Lambda,{\bf k})^{\dag}
\left(
n_1+n_2
\right)
U(\Lambda,{\bf k})
=
n_1+n_2,
\\
&&\hspace{-6pt}U(\Lambda,{\bf k})^{\dag}
\left(
n_3-n_0
\right)
U(\Lambda,{\bf k})
=n_3-n_0.
\end{eqnarray}
This implies that the Lorentz transformation alone does not mix the two observable in experiments transverse polarizations with the two unobservable ones.
\subsection{Four-translations in four-dimensional oscillator representation}\label{sec:P}
Let us begin  with the $N=1$ oscillator representation and denote $U({\bf 1},y,1)= e^{iP(1)\cdot y}$.
The generator of four-translations, the four-momentum for the reducible representation reads
\begin{eqnarray}\label{P(1)}
P_{\bf{a}}(1)
&=&
{\textstyle\int} d\Gamma({\bf k})k_{\bf{a}}|{\bf k}\rangle
\langle {\bf k}|\otimes H,
\end{eqnarray}
where $P_0(1)$ is of course the covariant Hamiltonian (\ref{covredH}) introduced earlier in section \ref{sec:4dreducible}.
One can immediately verify that
\begin{eqnarray}
e^{iP(1)\cdot y}a_{\bf a}({\bf k},1)e^{-iP(1)\cdot y}
&=&
a_{\bf a}({\bf k},1) e^{-iy\cdot k},
\end{eqnarray}
implying the following transformation on the vector potential
\begin{eqnarray}
U({\bf 1},y,1)^{\dagger}A_{a}(x,1)U({\bf 1},y,1)
&=&
A_{a}(x-y,1).
\end{eqnarray}
Furthermore, the four-momentum for arbitrary \linebreak $N$-oscillator reads
\begin{eqnarray}\label{P(N)}
P_{\bf{a}}(N)
&=&
\sum_{n=1}^N P_{\bf{a}}(1)^{(n)}
.
\end{eqnarray}
Vectors (\ref{ket41N}) are simultaneously the eigenvectors \linebreak of $P_{\bf{a}}(N)$, i.e.
\begin{eqnarray}
&&\hspace{-6pt}P^{\bf{a}}(N) |{\bf k}_1,\dots,{\bf k}_N,
n_0^{1},\dots,n_3^{N}\rangle
\nonumber\\
&&=
\Big(k^{\bf{a}}_1\big(n_1^{1}+n_2^{1}+n_3^{1}-n_0^{1}+1\big)
+\dots
\nonumber\\
&&\quad
+~k^{\bf{a}}_N\big(n_1^{N}+n_2^{N}+n_3^{N}-n_0^{N}+1\big)\Big)
\nonumber\\
&&\quad\times~|{\bf k}_1,\dots,{\bf k}_N, n_0^{1},\dots,n_3^{N}\rangle.
\end{eqnarray}
Then the following transformation rule for the ladder operators in the $N$-oscillator reducible representation holds
\begin{eqnarray}
e^{iP(N)\cdot y}a_{\bf a}({\bf k},N)e^{-iP(N)\cdot y}
&=&
a_{\bf a}({\bf k},N) e^{-iy\cdot k},
\end{eqnarray}
implying
\begin{eqnarray}
U({\bf 1},y,N)^{\dagger}A_{a}(x,N)U({\bf 1},y,N) &=& A_{a}(x-y,N).\quad
\end{eqnarray}
Then the Poincar\'e group, i.e. the semi-direct product of homogeneous Lorentz transformation and space-time translation groups is
\begin{eqnarray}
U(\Lambda,y,1)
~=~
U({\bf 1},y,1) U(\Lambda,0,1),
\end{eqnarray}
and the composition law of two successive Poincar\'e transformations holds
\begin{eqnarray}
U(\Lambda_2,y_2,1) U(\Lambda_1,y_1,1)
~=~
U(\Lambda_2\Lambda_1,\Lambda_2y_1+y_2,1).\qquad
\end{eqnarray}
\section{Summary and conclusions}\label{sec:4dredconclusions}
In summary, we have proposed a construction for the four-dimensional polarization space coming from a definition of a covariant Hamiltonian (\ref{covH}).  Further analysis is presented for such formalism, especially regarding the interpretation of the ladder operators for the time-like polarization. Strong arguments are given in favor of an interpretation in which the operator annihilating vacuum is a raising energy operator. It turns out that assuming for the time-like photons the energy spectrum bounded from the top, we can preserve the positive norms as needed for the probability interpretation of quantum mechanics. Such an interpretation gives a non divergent vacuum representation and saves the hermiticity operation. Analysis presented in this paper is in agreement with the four-dimensional quantization of the potential operator in mentioned \cite{MCJN05}, \cite{MCKW09} papers. Further reducible representation algebras for the four-dimensional polarization oscillators are presented. Using the covariant Hamiltonian (\ref{covredHN}) for $N$-oscillator reducible representation algebras, we find that such formalism is free from vacuum energy divergences. We show that the convergence of vacuum is guaranteed by a proper choice of the vacuum probability density function $Z({\bf k})$.
Next, states $\Psi_{EM}$, which reproduce standard electromagnetic fields (i.e. photons with two transverse polarizations) from the four-dimensional covariant formalism (i.e. with two additional longitudinal and time-like photons) are shown explicitly and discussed. It is interesting that such states have a coherent-like structure. 
Finally, a homogeneous Lorentz transformation for the four-dimensional oscillator is introduced.  Further, generators of these transformations coming from the canonical variables are shown.
We point out that there exists a transformation that corresponds to a gauge transformation of the potential for $\Psi_{EM}$ vectors. Next, invariants of introduced transformations are shown. Let us stress that the ``ghost operator" coming from the two extra degrees of photon polarization is an invariant. 
\begin{acknowledgements}
	I am grateful to Marek Czachor and Gosia \'Smia\l{}ek-Telega for discussions and critical comments.
\end{acknowledgements}
\appendix
\begin{small}
\section{Penrose abstract indices notation for tetrads}\label{sec:AppA}
Three types of indices were introduced in this paper: the boldface indices
${\bf{a}}$, ${\bf{b}}$ take numerical values $0$, $1$, $2$, $3$ and are related to a concrete choice of basis in Minkowski tetrad. The italics $a$, $b$ are abstract indices and specify types of tensor objects.
Indices that are boldfaced primed ${{\bf a}'}=00',01',10',11'$ are related to a concrete choice of basis in null tetrad. 
The abstract index formalism allows to work at a basis independent level, with all the operations on indices we know from the matrix calculus.

However first we will introduce the Penrose spin-frame notation. The symplectic form $\varepsilon_{AB}$ is a skew-symmetric complex bilinear form, i.e. $\varepsilon_{AB}=-\varepsilon_{BA}$, 
such that the action of the bilinear form on vectors $\varepsilon_{AB}\psi^A\phi^B$ is a complex number. The element of the dual spin-space is written with a unprimed subscript $\psi_A$. 
It is often convenient to use a collective symbol $\varepsilon{_{{\bf A}}}{^A}$ for a spin-frame basis such that
\begin{equation}
\varepsilon_{0}{^{A}}=\omega^{A},
\qquad
\varepsilon_{1}{^{A}}=\pi^{A}.
\end{equation}
Then the components of $\varepsilon_{AB}$ with respect to the spin-frame basis are
\begin{eqnarray}
\varepsilon_{{\bf A} {\bf {B}}}
=
\varepsilon_{A B}
\varepsilon_{{\bf A}}{^A}
\varepsilon_{{\bf {B}}}{^B}
=
\left(
\begin{array}{cc}
0&1\\
-1&0
\end{array}
\right).
\end{eqnarray}
The dual basis denoted as $\varepsilon{_A}{^{{\bf A}}}$ must then satisfy
\begin{eqnarray}
\varepsilon_{{\bf A}}{^A}
\varepsilon_{A}{^{{\bf {B}}}}
=
\varepsilon_{{\bf A}}{^{ {\bf {B}}}}
=
\left(
\begin{array}{cc}
1&0\\
0&1
\end{array}
\right).
\end{eqnarray}
From this it implies that
\begin{equation}
\varepsilon_{A}{^0}=-\pi_A,~~~~
\varepsilon_{A}{^1}=\omega_A.
\end{equation}
Therefore, the spin-frame and dual spin-frame can be written in a matrix notation, i.e.
\begin{eqnarray}
\varepsilon_{ A}{^{{\bf A}}}\label{epsilon^01}
&=&
\left(
\begin{array}{cc}
-\pi_A \\
\omega_A
\end{array}
\right)
,
\qquad
\varepsilon_{{\bf A}}{^{ A}}  
~=
\left(
\begin{array}{cc}
\omega^A\\
\pi^A
\end{array}
\right),
\end{eqnarray}
and any spin-frame satisfies the following
\begin{eqnarray}
\varepsilon_{AB}=\omega_{A}\pi_{B}-\pi_{A}\omega_{B}
\end{eqnarray}
Returning to the tetrads, the Minkowski space has the signature $(+,-,-,-)$ and the metric tensor is denoted by $g_{ab}$, $g^{ab}$. $g_{{\bf {ab}}}$ and $g^{{\bf {ab}}}$ are matrices diag$(+,-,-,-)$.
Minkowski tetrad $g_{a}{^{\bf{a}}}$, indexed by indices that are partly boldfaced and partly italic, consists of four four-vectors $g_{a}{^0}$, $g_{a}{^1}$, $g_{a}{^2}$, $g_{a}^{~3}$. There are two types of tetrads, related to the four-momentum $k^{{\bf{a}}}$ of a massless particles. The momentum independent tetrad $g_{a}^{~{\bf a}}$ satisfies $k^0=|{\bf k}|=k^a g_{a}{^0}$, $k^1=k^a g_{a}{^1}$, $k^2=k^a g_{a}{^2}$, $k^3=k^a g_{a}{^3}$, and defines decomposition into energy and momentum in the Lorentz invariant measure.
The null-vector $k^{a}=k^{a}({\bf k})$ plays the role of a flag pole for spinor field $\pi^{A}({\bf k})$ and can be written
in a spinor notation:
$k^{a}({\bf k})=\pi^{A}({\bf k})\pi^{A'}({\bf k})$,
where $\pi^{A}({\bf k})$ is a spinor field defined by $k^{a}({\bf k})$
up to a phase factor. For any $\pi^{A}({\bf k})$ there is
another spinor field associated $\omega^{A}({\bf k})$
satisfying the spin-frame condition $\omega_{A}({\bf k})\pi^{A}({\bf k})=1$.

We also consider a general field of tetrads defined on the light cone $g_{a}^{~~{\bf a}}({\bf k})$. Here $g_{a}^{~~1}({\bf k})$ and  $g_{a}^{~~2}({\bf k})$ can play role of transverse polarization vectors.
Indices $a$ and ${\bf a}$ can be raised and lowered by means of the
Minkowski metric tensor $g_{ab}$, $g^{ab}$, $g_{\bf{ab}}$ and $g^{\bf{ab}}$.

The null tetrad is indexed by indices that are partly boldfaced primed and partly italic
$g^{a}_{~~{\bf{b}}'}$
It is important to distinguish between ${\bf a}$ and $\bf{a'}$, and we will employ the convention where ${{\bf a}'}=00',01',10',11'$ and $g_{\bf{a'b'}}=\varepsilon_{\bf{AB}}\varepsilon_{\bf{A'B'}}$, 
$g^{\bf{a'b'}}=\varepsilon^{\bf{AB}}\varepsilon^{\bf{ A'B'}}$, where ${{\bf A}}=0,1$ and  ${{\bf A'}}=0',1'$. We raise and lower indices  
$\bf{AA'}$ by means of $\varepsilon_{\bf{AB}}\varepsilon_{\bf{A'B'}}$ and indices ${\bf a}'$ by means of matrix
\begin{equation}
g_{\bf{a'b'}}\label{gdoa'b'}
=
\left(
\begin{array}{cccc}
0&0&0&1\\
0&0&-1&0\\
0&-1&0&0\\
1&0&0&0\\
\end{array}
\right).
\end{equation}
Therefore the null tetrad associated with spin-frames can be written as
\begin{eqnarray}\label{nulltetrad}
g^{a}_{~~{\bf{b}}'}
&=&
\varepsilon^{A}{_{\bf B}}
\varepsilon^{A'}{_{\bf B'}}
=
\left(
\begin{array}{c}
\omega^{A}\omega^{A'}\\
\omega^{A}\pi^{A'}\\
\pi^{A}\omega^{A'}\\
\pi^{A}\pi^{A'}
\end{array}
\right)
=
\left(
\begin{array}{c}
\omega^{a}\\
m^{a}\\
\bar{m}^{a}\\
k^{a}
\end{array}
\right),\qquad
\nonumber\\
g_{a}^{~~{\bf{b}}'}
&=&
\varepsilon_{A}{^{\bf B}}
\varepsilon_{A'}{^{\bf B'}}
=
\left(
\begin{array}{c}
\pi_{A}\pi_{A'}\\
-\pi_{A}\omega_{A'}\\
-\omega_{A}\pi_{A'}\\
\omega_{A}\omega_{A'}
\end{array}
\right)
=
\left(
\begin{array}{c}
k_{a}\\
-\bar{m}_{a}\\
-m_{a}\\
\omega_{a}
\end{array}
\right).\qquad
\end{eqnarray}
Here, $g^{a}{_{01'}}$ and $g^{a}{_{10'}}$ can play the role for circular photon polarization vectors.

There is a relation between Minkowski tetrad, indexed by indices that are
partly boldfaced and partly italic, and null tetrad
\begin{eqnarray}
g^{a}{_{\bf{a}}}
=
g_{{\bf{a}}}^{~~{\bf{b}}'}
g^{a}_{~~{\bf{b}}'}
=
g_{{\bf{a}}}^{~~\bf{BB}'}
g^{a}_{~~\bf{BB}'},
\end{eqnarray}
\begin{eqnarray}\label{Mintetrad}
g^{a}{_{\bf{a}}} 
&=&
\left(
\begin{array}{c}
g^{a}_{~~0}\\
g^{a}_{~~1}\\
g^{a}_{~~2}\\
g^{a}_{~~3}
\end{array}
\right) 
= 
\frac{1}{\sqrt{2}} \left(
\begin{array}{cccc}
1&0&0&1\\
0&1&1&0\\
0&i&-i&0\\
1&0&0&-1
\end{array}
\right) \left(
\begin{array}{c}
\omega^{A}\omega^{A'}\\
\omega^{A}\pi^{A'}\\
\pi^{A}\omega^{A'}\\
\pi^{A}\pi^{A'}
\end{array}
\right)
\nonumber\\
&=&
\frac{1}{\sqrt{2}}
\left(
\begin{array}{c}
\omega^{a}+k^{a}\\
m^{a}+\bar{m}^{a}\\
i m^{a} - i\bar{m}^{a}\\
\omega^{a}-k^{a}
\end{array}
\right)
=
\left(
\begin{array}{c}
t^{a}\\
x^{a}\\
y^{a}\\
z^{a}
\end{array}
\right),
\end{eqnarray}
and dually we can write
\begin{eqnarray}
g_{a}{^{\bf{a}}}
=
g^{{\bf{a}}}_{~~{\bf{b}}'}
g_{a}^{~~{\bf{b}}'}
=
g^{{\bf{a}}}_{~~\bf{BB}'}
g_{a}^{~~\bf{BB}'},
\end{eqnarray}
\begin{eqnarray}
g_{a}{^{\bf{a}}}
&=&
\left(
\begin{array}{c}
g_{a}^{~~0}\\
g_{a}^{~~1}\\
g_{a}^{~~2}\\
g_{a}^{~~3}
\end{array}
\right) 
=
\frac{1}{\sqrt{2}} \left(
\begin{array}{cccc}
1&0&0&1\\
0&1&1&0\\
0&-i&i&0\\
1&0&0&-1
\end{array}
\right) \left(
\begin{array}{c}
\pi_{A}\pi_{A'}\\
-\pi_{A}\omega_{A'}\\
-\omega_{A}\pi_{A'}\\
\omega_{A}\omega_{A'}
\end{array}
\right)
\nonumber\\
&=&
\frac{1}{\sqrt{2}}
\left(
\begin{array}{c}
k_{a}+\omega_{a}\\
-\bar{m}_{a}-m_{a}\\
i \bar{m}_{a} -i m_{a}\\
k_{a}-\omega_{a}
\end{array}
\right)
=
\left(
\begin{array}{c}
t_{a}\\
-x_{a}\\
-y_{a}\\
-z_{a}
\end{array}
\right).
\end{eqnarray}
Here the $g$s with the partly boldfaced and partly boldfaced primed
indices are the Infeld-van der Waerden symbols which can be written in the following matrix forms
\begin{eqnarray}
g_{~~{\bf{b}}'}^{{\bf{a}}} \label{gupadob'}
=
\frac{1}{\sqrt{2}} \left(
\begin{array}{cccc}
1&0&0&1\\
0&1&1&0\\
0&-i&i&0\\
1&0&0&-1\\
\end{array}
\right), 
\end{eqnarray}
\\
\begin{eqnarray}
g_{{\bf{a}}}^{~~{\bf{b}}'} \label{gdoaupb'}
=
\frac{1}{\sqrt{2}} \left(
\begin{array}{cccc}
1&0&0&1\\
0&1&1&0\\
0&i&-i&0\\
1&0&0&-1\\
\end{array}
\right).
\end{eqnarray}
These Infeld-van der Waerden symbols in their matrix form, in Penrose abstract index formalism, are used to translate formulas into matrix forms. 
Furthermore, the relation between Minkowski tetrad and the metric tensor $g_{ab}$ is
\begin{eqnarray}\label{metrictensorM}
&&\hspace{-6pt}g_{ab}
=
g_a{^{{\bf a}}} ({\bf k})g_b{^{{\bf {b}}}} ({\bf k})
g_{{\bf a}{\bf {b}}}
\nonumber\\
&&\quad=
-x_{a}({\bf k})x_{b}({\bf k})
-y_{a}({\bf k})y_{b}({\bf k})
-z_{a}({\bf k})z_{b}({\bf k})
+t_{a}({\bf k})t_{b}({\bf k}).
\nonumber\\
\end{eqnarray}
Analogically we get the following relation between null tetrad and the metric tensor $g_{ab}$
\begin{eqnarray}\label{metrictensorN}
&&\hspace{-6pt}g_{ab}
=
g_a{^{{\bf a}'}} ({\bf k})g_b{^{{\bf {b}}'}} ({\bf k})
g_{{\bf a}'{\bf {b}}'}
\nonumber\\
&&=
k_{a}({\bf k})\omega_{b}({\bf k})
+\omega_{a}({\bf k})k_{b}({\bf k})
-m_{a}({\bf k})\bar{m}_{b}({\bf k})
-\bar{m}_{a}({\bf k})m_{b}({\bf k}).
\nonumber\\
\end{eqnarray}
\section{Transformation properties of spin-frames}\label{sec:AppB}
We introduce two symmetries that leave the spin-frame condition
\begin{equation}\label{s-fcondition}
\omega_{A}({\bf k})\pi^{A}({\bf k})=1
\end{equation}
invariant. First, the spinor field transformation associated with the homogeneous Lorentz transformation
\begin{eqnarray}
&&\hspace{-6pt}\pi_{A}({\bf k})~~\mapsto~~ \Lambda\pi_{A}({\bf k})
=
\Lambda_{A}^{~~B}\pi_{B}({\bf{\Lambda^{-1}k}}),
\nonumber\\
&&\hspace{-6pt}\pi^{A}({\bf k})~~\mapsto~~ \Lambda\pi^{A}({\bf k})
=
\pi^{B}({\bf{\Lambda^{-1}k}})
\Lambda^{-1}{_{B}}{^{A}}.
\end{eqnarray}
Here $\Lambda_{A}^{~~B}$ is
an unprimed SL(2,C) matrix corresponding to $\Lambda_{a}^{~~b}\in$~SO(1,3).
The null vector $k_a({\bf k})$ that plays the role of a flag-pole for the spinor field $\pi_A({\bf k})$, i.e. 
\begin{equation}k_a({\bf k})=\pi_{A}({\bf k})\pi_{A'}({\bf k}),
\end{equation}
must be invariant, so
\begin{equation}
\Lambda\pi_{A}({\bf k})
\Lambda\pi_{A'}({\bf k})
=\pi_{A}({\bf k})\pi_{A'}({\bf k})
\end{equation}
must be satisfied and hence
\begin{equation}\label{lorpitrans}
\Lambda\pi_{A}({\bf k})
=
e^{-i\Theta(\Lambda,{\bf k})}
\pi_{A}({\bf k}).
\end{equation}
The angle $\Theta(\Lambda,{\bf k})$ is called the Wigner phase. Note that in most literature it is the doubled value $2\Theta(\Lambda,{\bf k})$ which is called the Wigner phase. In analogy
\begin{eqnarray}
&&\hspace{-6pt}\omega_{A}({\bf k})~~\mapsto~~ \Lambda\omega_{A}({\bf k})
=
\Lambda_{A}^{~~B}\omega_{B}({\bf{\Lambda^{-1}k}}),
\nonumber\\
&&\hspace{-6pt}\omega^{A}({\bf k})~~\mapsto~~ \Lambda\omega^{A}({\bf k})
=
\omega^{B}({\bf{\Lambda^{-1}k}})
\Lambda^{-1}{_{B}}{^{A}}.
\end{eqnarray}
and the spin-frame condition (\ref{s-fcondition}) has to hold. We assume a special case, i.e.
\begin{equation}\label{loromtrans}
\Lambda\omega_{A}({\bf k})
=
e^{i\Theta(\Lambda,{\bf k})}
\omega_{A}({\bf k}).
\end{equation}

Let us also define another symmetry
\begin{eqnarray}\label{gaomtrans}
\omega_{A}({\bf k})~~&\mapsto&~~
\tilde{\omega}_{A}({\bf k})
=
\omega_{A}({\bf k})+\phi({\bf k})\pi_{A}({\bf k}),
\\
\pi_{A}({\bf k})~~&\mapsto&~~\label{gapitrans}
\tilde{\pi}_{A}({\bf k})
=\pi_{A}({\bf k}),
\end{eqnarray}
which also keeps the spin-frame condition (\ref{s-fcondition}). Here $\phi({\bf k}) = |\phi({\bf k})| e^{i\xi({\bf k})}$
is at this point any complex number. It is interesting that the ambiguity of $\phi({\bf k})$ at the spinor level manifests itself in electrodynamics.
The spin-frame condition
\begin{equation}\label{spinfrcond}
\Lambda\tilde{\omega}_{A}({\bf k})\Lambda\pi^{A}({\bf k})
= 1
\end{equation}
holds for the most general transformation  written as
\begin{eqnarray}\label{omegatildelambda}
\Lambda\tilde{\omega}_{A}({\bf k})
&=&
e^{i\Theta(\Lambda,{\bf k})}
\left(\omega_{A}({\bf k})+\phi({\bf k})\pi_{A}({\bf k})\right)
=
e^{i\Theta(\Lambda,{\bf k})}\tilde{\omega}_{A}({\bf k}).
\nonumber\\
\end{eqnarray}
Also
\begin{eqnarray}
&&\hspace{-6pt}\Lambda\tilde{\omega}_{A}({\bf k})
=
\Lambda_{A}^{~~B}\omega_{B}({\bf{\Lambda^{-1}k}})
+
\phi({\bf{\Lambda^{-1}k}})\Lambda_{A}^{~~B}\pi_{B}({\bf{\Lambda^{-1}k}})\qquad
\nonumber\\
&&	=
e^{i\Theta(\Lambda,{\bf k})}
\left(\omega_{A}({\bf k})+
\phi({\bf{\Lambda^{-1}k}})e^{-2i\Theta(\Lambda,{\bf k})}\pi_{A}({\bf k})\right),
\end{eqnarray}
and from this the Lorentz transformation rule for $\phi({\bf k})$ is
\begin{equation}\label{transphi}
\Lambda\phi({\bf k})
=
\phi({\bf{\Lambda^{-1}k}})
=
e^{2i\Theta(\Lambda,{\bf k})}
\phi({\bf k}).
\end{equation}
It should be stressed that there is a difference in interpretation of $\phi({\bf k})$ compared with \cite{MCJN05} and \cite{MCKW09}. Here the Lorentz transformation is not parametrized by $\phi({\bf k})$ and this implies a difference in (\ref{omegatildelambda}) notation, when compared with $\Lambda\omega_{A}({\bf k})
=
e^{i\Theta(\Lambda,{\bf k})}
\left(\omega_{A}({\bf k})+\phi({\bf k})\pi_{A}({\bf k})\right)$ from \cite{MCJN05} and \cite{MCKW09}.  This way the succession of these two transformations is emphasised. It is important to stress that the Lorentz transformation and the transformation parametrized by $\phi({\bf k})$ do not commute.

Now let us introduce an SL(2,C) matrix that links two spin-frames
\begin{eqnarray}\label{LAB}
L_{\bf{A}}^{~~{\bf{B}}}(\Theta,\phi)
=
\varepsilon_{\bf{A}}^{~~A}({\bf k})
\Lambda
\tilde{\varepsilon}^{~~{\bf{B}}}_{A}({\bf k}).
\end{eqnarray}
This matrix can explicitly be written in terms of the Wigner phase  $\Theta(\Lambda, {\bf k})$ and the $\phi({\bf k})$ parameter, i.e.
\begin{eqnarray}\label{LABmatrix}
L_{\bf{A}}^{~~{\bf{B}}}(\Theta,\phi)
&=&
\left(
\begin{array}{cc}
\omega_{A}({\bf k})
\Lambda
\pi^{A}({\bf k})&
\omega^{A}({\bf k})
\Lambda
\tilde{\omega}_{A}({\bf k})\\
0&
\pi^{A}({\bf k})
\Lambda
\tilde{\omega}_{A}({\bf k})
\end{array}
\right)
\nonumber\\
&=&
\left(
\begin{array}{cc}
e^{-i\Theta(\Lambda,{\bf k})}
& -\phi({\bf k})e^{i\Theta(\Lambda,{\bf k})}\\
0 & e^{i\Theta(\Lambda,{\bf k})}\\
\end{array}
\right).	
\end{eqnarray}
Furthermore, this matrix can be split into two SL(2,C) matrices corresponding to gauge and homogeneous Lorentz transformations, i.e.
\begin{eqnarray}\label{LABex}
L_{\bf A}{^{\bf B} }(\Theta,\phi)
&=&
\left(
\begin{array}{cc}
1
& -\phi({\bf k})\\
0 &1\\
\end{array}
\right)
\left(
\begin{array}{cc}
e^{-i\Theta(\Lambda,{\bf k})}
&0\\
0 & e^{i\Theta(\Lambda,{\bf k})}\\
\end{array}
\right)
.\quad\quad
\end{eqnarray}
Finally, let us define those two matrices corresponding to gauge transformation and Wigner rotations respectively
\begin{eqnarray}\label{GABex}
G_{\bf A}{^{\bf B}}
({\bf k})
&=&
L_{\bf A}{^{\bf B} }(0,\phi({\bf k}))
=
\varepsilon_{\bf A}^{~~A}({\bf k})
\tilde{\varepsilon}^{~~{\bf B}}_{A}({\bf k})
\nonumber\\
&=&
\left(
\begin{array}{cc}
1
& -\phi({\bf k})\\
0 &1\\
\end{array}
\right),
\end{eqnarray}
\begin{eqnarray}\label{RAB}
R_{\bf A}{^{\bf B} }(\Lambda,{\bf k})
&=&
L_{\bf A}{^{\bf B} }(\Theta(\Lambda,{\bf k}),0)
=
\varepsilon_{\bf A}^{~~A}({\bf k})
\Lambda
\varepsilon^{~~{\bf B}}_{A}({\bf k})
\label{RABex}
\nonumber\\
&=&
\left(
\begin{array}{cc}
e^{-i\Theta(\Lambda,{\bf k})}
&0\\
0 & e^{i\Theta(\Lambda,{\bf k})}\\
\end{array}
\right).
\end{eqnarray}
\section{Transformation properties of tetrads}\label{sec:AppC}
In this appendix we will introduce a transformation on the Minkowski and null tetrads that corresponds to the spin-frame transformations (\ref{lorpitrans}), (\ref{loromtrans}),   (\ref{gaomtrans}) and (\ref{gapitrans}).
Now, for any homogeneous Lorentz transformation $\Lambda_{a}{^{b}}$, parametrized by the Wigner phase $\Theta(\Lambda, {\bf k})$, and any gauge transformation parametrized by some complex number $\phi({\bf k})$, let us define  the following
matrix associated with the Minkowski tetrad
\begin{equation}\label{Lab}
L_{{\bf{a}}}{^{\bf{b}}}
(\Theta,\phi)
=
g_{~~{\bf{a}}}^{a}({\bf k})
\Lambda_{a}^{~~b}
\tilde{g}^{~~{\bf{b}}}_{b}
({\bf{\Lambda^{-1}k}})
=
g_{~~{\bf{a}}}^{a}
({\bf k})\Lambda
\tilde{g}^{~~{\bf{b}}}_{a}
({\bf k}).
\end{equation}
Matrix $L_{\bf{a}}{^{\bf {b}}}$ can be written explicitly parametrized by
$\Theta(\Lambda,{\bf k})$ and $\phi({\bf k})$.
\begin{eqnarray}\label{Labap}
&&\hspace{-6pt}L_{{\bf{a}}}{^{\bf{b}}}(\Theta,\phi)
\nonumber\\
&&=
\left(
\begin{array}{cccc}
1+\frac{|\phi|^2}{2} 
& -|\phi|\cos(\xi+2\Theta) 
&|\phi|\sin(\xi+2\Theta) 
&-\frac{|\phi|^2}{2}
\\
-|\phi|\cos \xi
& \cos 2\Theta
& -\sin 2\Theta
&|\phi|\cos \xi
\\
|\phi|\sin \xi
&\sin 2\Theta
& \cos 2\Theta
&-|\phi|\sin \xi
\\
\frac{|\phi|^{2}}{2}
& -|\phi|\cos(\xi+2\Theta)
& |\phi|\sin(\xi+2\Theta) 
& 1-\frac{|\phi|^{2}}{2}
\end{array}
\right).\label{LLLLL}
\nonumber\\
\end{eqnarray}
Here $\Theta=\Theta(\Lambda,\bf k)$, $\phi=\phi(\bf k)$ and $\xi=\xi(\bf k)$.
This matrix has the property of linking two Minkowski tetrads in a way that
\begin{eqnarray}
g_{c}{^{\bf{a}}}({\bf k})
L_{{\bf{a}}}{^{\bf{b}}}
(\Theta,\phi)
&=&
g_{c}{^{\bf{a}}}({\bf k})
g_{~~{\bf{a}}}^{a}({\bf k})
\Lambda_{a}{^b}
\tilde{g}^{~~{\bf{b}}}_{b}
({\bf{\Lambda^{-1}k}})
\nonumber\\
&=&
g_{c}^{~~a}
\Lambda_{a}^{~~b}
\tilde{g}^{~~{\bf{b}}}_{b}
({\bf{\Lambda^{-1}k}})
\nonumber\\
&=&
\Lambda_{c}^{~~b}
\tilde{g}^{~~{\bf{b}}}_{b}
({\bf{\Lambda^{-1}k}}).
\end{eqnarray}
The metric tensor must be invariant under this combined Lorentz and gauge transformation, therefore
\begin{eqnarray}
&&\hspace{-6pt}g_{a}{^{\bf{a}}}({\bf k})
g_{b}{^{\bf{b}}}({\bf k})
g_{\bf{ab}}
\nonumber\\
&&=
g_{a}{^{\bf{c}}}({\bf k})
L_{\bf{c}}{^{\bf{a}}}(\Theta,\phi)
g_{b}{^{\bf{d}}}({\bf k})
L_{\bf{d}}{^{\bf{b}}}(\Theta,\phi)
g_{\bf{ab}}
\nonumber\\
&&=
g_{a}{^{\bf{c}}}({\bf k})
g_{b}{^{\bf{d}}}({\bf k})
L_{\bf{c}}{^{\bf{a}}}(\Theta,\phi)
L_{{\bf{d}}{\bf{a}}}(\Theta,\phi)
\end{eqnarray}
and this implies 
\begin{eqnarray}
&&\hspace{-6pt}
L_{{\bf{a}}}{^{\bf{b}}}
(\Theta,\phi)
L_{\bf{cb}}
(\Theta,\phi)
=
L_{\bf{ab}}
(\Theta,\phi)
L_{\bf{c}}{^{\bf{b}}}
(\Theta,\phi)
=
g_{\bf{ac}},\qquad
\\
&&\hspace{-6pt}
L_{{\bf{a}}}{^{\bf{b}}}
(\Theta,\phi)
L^{\bf{c}}{_{\bf{b}}}
(\Theta,\phi)
=
L_{\bf{ab}}
(\Theta,\phi)
L^{\bf{cb}}
(\Theta,\phi)
=
g_{{\bf{a}}}{^{\bf{c}}},
\\
&&\hspace{-6pt}L^{-1}{_{\bf{a}}}{^{\bf{b}}}\label{L-1toLT}
(\Theta,\phi)
=
L{^{\bf{b}}}{_{\bf{a}}}
(\Theta,\phi)
.
\end{eqnarray}
$L_{{\bf{a}}}{^{\bf{b}}}(\Theta,\phi)$ can be written as a product of the Wigner rotation matrix $R_{{\bf{a}}}{^{\bf{b}}}(\Lambda,{\bf k}) $ and a gauge transformation matrix $G_{{\bf{a}}}{^{\bf{b}}}({\bf k})$, i.e.
\begin{eqnarray}
&&\hspace{-6pt}
L_{{\bf{a}}}{^{\bf{b}}}(\Theta,\phi)
=
G_{{\bf{a}}}{^{\bf{c}}}({\bf k})
R_{\bf{c}}{^{\bf{b}}}(\Lambda,{\bf k}),
\end{eqnarray}
where
\begin{eqnarray}\label{Gab}
G_{{\bf{a}}}{^{\bf{b}}}({\bf k})
&=&
\left(
\begin{array}{cccc}
1+\frac{|\phi|^2}{2}
& -|\phi|\cos\xi
&|\phi|
\sin\xi
&-\frac{|\phi|^2}{2}
\\
-|\phi|\cos \xi
& 1
& 0
&|\phi|\cos \xi
\\
|\phi|\sin \xi
&0
& 1
&
-|\phi|\sin \xi
\\
\frac{|\phi|^{2}}{2}
& -|\phi|
\cos\xi
& |\phi|
\sin\xi
& 1-\frac{|\phi|^{2}}{2}
\end{array}
\right),
\end{eqnarray}
\begin{eqnarray}\label{Rab}
R_{\bf{a}}{^{\bf{b}}}(\Lambda,{\bf k})=
\left(
\begin{array}{cccc}
1
&0
&0
&0
\\
0
& \cos 2\Theta
& -\sin 2\Theta
&0
\\
0
&\sin 2\Theta
& \cos 2\Theta
&0
\\
0
& 0
& 0
& 1
\end{array}
\right).
\end{eqnarray}
Of course, those two matrices do not commute, i.e.
\begin{eqnarray}\label{GRRG}
L_{{\bf{a}}}{^{\bf{b}}}(\Theta,\phi)
=
G_{{\bf{a}}}{^{\bf{c}}}({\bf k})
R_{\bf{c}}{^{\bf{b}}}(\Lambda,{\bf k})
&=&
R_{{\bf{a}}}{^{\bf{c}}}(\Lambda,{\bf k})
G_{\bf{c}}{^{\bf{b}}}({\bf{\Lambda^{-1}k}}).
\nonumber\\
\end{eqnarray}
From (\ref{Lab}) we can get
\begin{eqnarray}
L_{{\bf{a}}}{^{\bf{b}}}(\Theta,\phi)
&=&
g_{~~{\bf{a}}}^{a}({\bf k})
\Lambda_{a}^{~~b}
\tilde{g}^{~~{\bf{b}}}_{b}
({\bf{\Lambda^{-1}k}})
\nonumber\\
&=&
g_{{\bf{a}}}^{~~{\bf{a}}'}
g_{~~{\bf{a}}'}^{a}({\bf k})
g^{\bf{b}}_{~~{\bf{b}}'}
\Lambda_{a}^{~~b}
\tilde{g}_{b}^{~~{\bf{b}}'}
({\bf{\Lambda^{-1}k}})
\nonumber\\
&=&
g_{{\bf{a}}}^{~~{\bf{a}}'}
g^{\bf{b}}_{~~{\bf{b}}'}
L_{{\bf{a}}'}^{~~{\bf{b}}'}
(\Theta,\phi).
\end{eqnarray}
$g_{{\bf{a}}}{^{\bf{a}'}}$ and $g{_{\bf{b}'}}^{\bf{b}}$ are the Infeld-van der Waerden symbols introduced in (\ref{gupadob'})--(\ref{gdoaupb'}) and matrix $L_{{\bf{a}}'}^{~~{\bf{b}}'}$ links two null tetrads
\begin{eqnarray}
&&\hspace{-6pt}L_{{\bf{a}}'}{^{{\bf{b}}'}}(\Theta,\phi)
\nonumber\\
&&=
g_{~~{\bf{a}}'}^{a}({\bf k})
\Lambda_{a}^{~~b}
\tilde{g}^{~~{\bf{b}}'}_{b}
({\bf{\Lambda^{-1}k}})
\nonumber\\
&&=
\varepsilon_{\bf{A}}^{~~A}({\bf k})
\varepsilon_{{\bf{A}}'}^{~~A'}({\bf k})
\Lambda_{A}^{~~B}
\Lambda_{A'}^{~~B'}
\tilde{\varepsilon}^{~~{\bf{B}}}_{B}
({\bf{\Lambda^{-1}k}})
\tilde{\varepsilon}^{~~{\bf{B}}'}_{B'}
({\bf{\Lambda^{-1}k}})
\nonumber\\
&&=
L_{\bf{A}}^{~~{\bf{B}}}(\Theta,\phi)
L_{{\bf{A}}'}^{~~{\bf{B}}'}(\Theta,\phi),
\end{eqnarray}
where $L_{\bf{A}}^{~~{\bf{B}}}(\Theta,\phi)$ is the transformation matrix (\ref{LAB}) linking two spin-frames.

\section{Composition law}\label{sec:AppD}
From (\ref{Wop}) and (\ref{Wopdag}) it is easy to show that $W(\Lambda)$ and $W(\Lambda)^{\dag}$ satisfy the following composition law
\begin{eqnarray}
&&\hspace{-6pt}W(\Lambda)
W(\Lambda')
|{\bf {k}}\rangle
=
|{\bf {\Lambda \Lambda' k}}\rangle
=
W(\Lambda \Lambda')
|{\bf k}\rangle ,
\\
&&\hspace{-6pt}W(\Lambda)^{\dag}
W(\Lambda')^{\dag}
|{\bf {k}}\rangle
=
|{\bf {(\Lambda'\Lambda)^{-1}k}}\rangle
=
W(\Lambda' \Lambda)^{\dag}
|{\bf k}\rangle
.\qquad
\end{eqnarray}
This means that they are unitary representations of the Lorentz group. We shall now take a closer look at the composition law for the reducible representations of $U(\Lambda,0,1)$ (\ref{LorentzTransN1})
\begin{eqnarray}\label{Ulambdalambda'}
&&\hspace{-6pt}
U(\Lambda,0,1)U(\Lambda',0,1)
\nonumber\\
&&=
\left(
\int d\Gamma({\bf k})
|{\bf k}
\rangle\langle
{\bf{\Lambda^{-1}k}}|\otimes U(\Lambda,{\bf k})
\right)
\nonumber\\
&&~~\times
\left(
\int d\Gamma({\bf k}')
|{\bf k}'
\rangle\langle
{\bf{\Lambda'^{-1}k'}}|\otimes U(\Lambda',{\bf k}')
\right)
\nonumber\\
&&=
\int d\Gamma({\bf k})
|{\bf k}\rangle
\langle {\bf{(\Lambda\Lambda')^{-1}k}}|
\otimes U(\Lambda,{\bf k})
U(\Lambda',{\bf{\Lambda^{-1}k}}).\qquad
\end{eqnarray}
The left-hand side of the above should be equal to \\$\int d\Gamma({\bf k})|{\bf k}\rangle\langle {\bf{(\Lambda\Lambda')^{-1}k}}|\otimes U(\Lambda\Lambda',{\bf k})$ and this implies the following condition
\begin{eqnarray}\label{Ucomposition}
U(\Lambda\Lambda',{\bf k})
&=&
U(\Lambda,{\bf k})
U(\Lambda',{\bf{\Lambda^{-1}k}}).
\end{eqnarray}
For the hermitian conjugates we can write
\begin{eqnarray}
&&\hspace{-6pt}
U(\Lambda',0,1)^{\dag}U(\Lambda,0,1)^{\dag}
\nonumber\\
&&=
\left(
\int d\Gamma({\bf k}')
|{\bf{\Lambda'^{-1}k'}}\rangle
\langle {\bf k}'|
\otimes U(\Lambda',{\bf k}')^{\dag}
\right)
\nonumber\\
&&~~\times
\left(
\int d\Gamma({\bf k})
|{\bf{\Lambda^{-1}k}}\rangle
\langle {\bf k}|
\otimes U(\Lambda,{\bf k})^{\dag}
\right)
\nonumber\\
&&=
\int d\Gamma({\bf k})
|{\bf{(\Lambda\Lambda')^{-1}k}}\rangle
\langle {\bf k}|
\otimes
U(\Lambda',{\bf{\Lambda^{-1}k}})^{\dag}
U(\Lambda,{\bf k})^{\dag}
.\qquad
\end{eqnarray}
This implies
\begin{eqnarray}
U(\Lambda\Lambda',{\bf k})^{\dag}
&=&
U(\Lambda',{\bf{\Lambda^{-1}k}})^{\dag}
U(\Lambda,{\bf k})^{\dag}.
\end{eqnarray}
The reducible representation is unitary thus
\begin{eqnarray}
&&\hspace{-6pt}
U(\Lambda,0,1)U(\Lambda,0,1)^{\dag}
\nonumber\\
&&=
\left(
\int d\Gamma({\bf k})
|{\bf k}\rangle
\langle {\bf{\Lambda^{-1}k}}|
\otimes U(\Lambda,{\bf k})
\right)
\nonumber\\
&&\quad\times
\left(
\int d\Gamma({\bf k}')
|{\bf{\Lambda^{-1}k'}}\rangle
\langle {\bf k}'|
\otimes U(\Lambda,{\bf k}')^{\dag}
\right)
\nonumber\\
&&=
\int d\Gamma({\bf k})
|{\bf k}\rangle
\langle {\bf k}|
\otimes U(\Lambda,{\bf k})
U(\Lambda,{\bf k})^{\dag}
=
I
\otimes 1_4.
\end{eqnarray}
On the other hand from (\ref{Ulambdalambda'}) putting $\Lambda'=\Lambda^{-1}$ we get
\begin{eqnarray}
&&\hspace{-6pt}
U(\Lambda,0,1)U(\Lambda^{-1},0,1)
\nonumber\\
&&=
\left(
\int d\Gamma({\bf k})
|{\bf k}
\rangle\langle
{\bf{\Lambda^{-1}k}}|\otimes U(\Lambda,{\bf k})
\right)
\nonumber\\
&&\quad\times
\left(
\int d\Gamma({\bf k}')
|{\bf k}'
\rangle\langle
{\bf{\Lambda k'}}|\otimes U(\Lambda^{-1},{\bf k}')
\right)
\nonumber\\
&&=
\int d\Gamma({\bf k})
|{\bf k}\rangle
\langle {\bf k}|
\otimes U(\Lambda,{\bf k})
U(\Lambda^{-1},{\bf{\Lambda^{-1}k}})
=
I
\otimes 1_4,\qquad
\end{eqnarray}
and this implies
\begin{eqnarray}
U(\Lambda,{\bf k})^{\dag}
&=&
U(\Lambda^{-1},{\bf{\Lambda^{-1}k}}).
\end{eqnarray}
Condition (\ref{Ucomposition}) imposes a composition law for the  $R_{{\bf{a}}}^{~~{\bf{b}}}(\Lambda, {\bf k})$ matrix such that
\begin{eqnarray}\label{Labcomposition}
R_{{\bf{a}}}^{~~{\bf{b}}}(\Lambda\Lambda',{\bf k})
&=&
R_{{\bf{a}}}^{~~{\bf{c}}}(\Lambda,{\bf k})
R_{\bf{c}}^{~~{\bf{b}}}(\Lambda',{\bf{\Lambda^{-1}k}})
.
\end{eqnarray}
The composition law will be shown here on the spinor level of SL(2,C) matrix
\begin{eqnarray}\label{compRAB}
R_{\bf{A}}{^{\bf{C}}}(\Lambda,{\bf k})
R_{\bf{C}}{^{\bf{B}}}(\Lambda',{\bf{\Lambda^{-1}k}})
=
R_{\bf{A}}{^{\bf{B}}}(\Lambda\Lambda', {\bf k}).
\end{eqnarray}
{\bf Proof}: 
Let us first remind ourselves of formula (\ref{LAB}) and write the Wigner Rotation matrix
\begin{eqnarray}\label{RAB}
R_{\bf{A}}^{~~{\bf{B}}}(\Lambda, {\bf k})
=
\varepsilon_{\bf{A}}{^{A}}({\bf k})
\Lambda\varepsilon_{A}{^{\bf{B}}}({\bf k}).
\end{eqnarray}
We can show that
\begin{eqnarray}
&&\hspace{-6pt}R_{\bf{A}}{^{\bf{C}}}(\Lambda,{\bf k})
R_{\bf{C}}{^{\bf{B}}}(\Lambda',{\bf{\Lambda^{-1}k}})
\nonumber\\
&&=
\varepsilon_{\bf{A}}{^{B}}({\bf k})
\Lambda\varepsilon_{B}{^{\bf{C}}}({\bf k})
\varepsilon_{\bf{C}}{^{A}}({\bf{\Lambda^{-1}k}})
\Lambda'\varepsilon_{A}{^{\bf{B}}}({\bf{\Lambda^{-1}k}})
\nonumber\\
&&=
\varepsilon_{\bf{A}}{^{B}}({\bf k})
\Lambda\varepsilon_{B}{^{\bf{C}}}({\bf k})
\Lambda\varepsilon_{\bf{C}}{^{E}}({\bf{k}})\Lambda_{E}{^A}
\Lambda'_{A}{^{C}}
\varepsilon_{C}{^{\bf{B}}}({\bf{(\Lambda\Lambda')^{-1}k}})
\nonumber\\
&&=
\varepsilon_{\bf{A}}{^{B}}({\bf k})
\Lambda\varepsilon_{B}{^{\bf{C}}}({\bf k})
\Lambda\varepsilon_{\bf{C}}{^{E}}({\bf{k}})
(\Lambda\Lambda')_{E}{^C}
(\Lambda\Lambda')^{-1}{_{C}}{^{D}}
\Lambda\Lambda'\varepsilon_{D}{^{\bf{B}}}({\bf{k}})
\nonumber\\
&&=
\varepsilon_{\bf{A}}{^{B}}({\bf k})
\Lambda\varepsilon_{B}{^{\bf{C}}}({\bf k})
\Lambda\varepsilon_{\bf{C}}{^{A}}({\bf k})
\Lambda\Lambda'\varepsilon_{A}{^{\bf{B}}}({\bf k})
\nonumber\\
&&=
\varepsilon_{\bf{A}}{^{B}}({\bf k})
\varepsilon_{B}{^{A}}
\Lambda\Lambda'\varepsilon_{A}{^{\bf{B}}}({\bf k})
\nonumber\\
&&=
\varepsilon_{\bf{A}}{^{A}}({\bf k})
\Lambda\Lambda'\varepsilon_{A}{^{\bf{B}}}({\bf k})
=
R_{\bf{A}}{^{\bf{B}}}(\Lambda\Lambda', {\bf k}).
\end{eqnarray}
Furthermore, let us also prove the composition law of Lorentz transformations on the gauge parameter $\phi({\bf k})$, i.e.
\begin{equation}\label{}
\Lambda\Lambda'\phi({\bf k})
=
\phi({\bf{(\Lambda\Lambda')^{-1}k}})
=
e^{2i\Theta(\Lambda\Lambda',{\bf k})}
\phi({\bf k}).
\end{equation}
{\bf Proof}: First, let us denote
\begin{eqnarray}\label{}
\Lambda\Lambda'\phi({\bf k})
&=&
\Lambda'\phi({\bf{\Lambda^{-1}k}})
=
\phi({\bf{\Lambda'^{-1}(\Lambda^{-1}k}}))
\nonumber\\
&=&
\phi({\bf{(\Lambda\Lambda')^{-1}k}}).
\end{eqnarray}
Using such a notation and transformation rule (\ref{transphi}) for $\phi({\bf k})$, we can show that
\begin{eqnarray}
&&\hspace{-6pt}\phi({\bf{(\Lambda\Lambda')^{-1}k}})
\nonumber\\
&&=
\Lambda'\phi({\bf{\Lambda^{-1}k}})
=
e^{2i\Theta(\Lambda',{\bf{\Lambda^{-1}k}})}
\phi({\bf{\Lambda^{-1}k}})
\\
&&=
e^{2i\Theta(\Lambda',{\bf{\Lambda^{-1}k}})}
e^{2i\Theta(\Lambda,{\bf k})}
\phi({\bf k})
=
e^{2i\Theta(\Lambda\Lambda',{\bf k})}
\phi({\bf k}).
\qquad
\end{eqnarray}
The last step of this proof can be shown using composition low (\ref{compRAB}) in its explicit SL(2,C) matrix form (\ref{RAB}).
\section{Transformation properties of the potential and electromagnetic field operator}\label{sec:AppE}
Let us recall the potential operator for $N=1$ oscillator representation (\ref{potN1})
\begin{eqnarray}
A_{a}(x,1)
&=&
i\int d\Gamma({\bf k})
g_{a}{^{\bf a}}({\bf k})a_{{\bf a}}({\bf k},1) 
e^{-ik\cdot x}+\rm{H.c.}
\nonumber
\end{eqnarray}
The potential operator transforms under Lorentz transformation (\ref{WU}) as a four-vector
\begin{eqnarray}
U(\Lambda,0,1)^{\dag} A_{a}(x,1)U(\Lambda,0,1)
&=&
\Lambda{_a}{^b}A_{b}(\Lambda^{-1}x,1).\qquad
\end{eqnarray}
{\bf Proof}:
\begin{eqnarray}
&&\hspace{-6pt}
U(\Lambda,0,1)^{\dag} A_{a}(x,1)U(\Lambda,0,1)
\nonumber\\
&&=
i\int d\Gamma({\bf k})
g_{a}{^{\bf{a}}}({\bf k})
R_{\bf{a}}{^{\bf{b}}}(\Lambda,{\bf k})
a_{\bf{b}}({\bf{\Lambda^{-1}k}},1)
e^{-ik\cdot x}+\rm{H.c.}
\nonumber\\
&&=
i\int d\Gamma({\bf k})
\Lambda g_{a}{^{\bf{b}}}({\bf k})
a_{\bf{b}}({\bf{\Lambda^{-1}k}},1)
e^{-ik\cdot x}+\rm{H.c.}
\nonumber\\
&&=
i\int d\Gamma({\bf k})
\Lambda{_a}{^b} g_{b}{^{\bf{b}}}({\bf{\Lambda^{-1}k}})
a_{\bf{b}}({\bf{\Lambda^{-1}k}},1)
e^{-ik\cdot x}+\rm{H.c.}
\nonumber\\
&&=
i\int d\Gamma({\bf k})
\Lambda{_a}{^b} g_{b}{^{\bf{b}}}({\bf k})
a_{\bf{b}}({\bf k},1)
e^{-ik\cdot \Lambda^{-1}x}+\rm{H.c.}
\nonumber\\
&&=
\Lambda{_a}{^b}A_{b}(\Lambda^{-1}x,1).
~~~~
\end{eqnarray}
We can also extend this calculation to any natural number $N$, where the four-potential in $N$-oscillator representation is denoted by $A_a(x,N)$, i.e.
\begin{eqnarray}
U(\Lambda,0,N)^{\dag} A_{a}(x,N)U(\Lambda,0,N)
&=&
\Lambda{_a}{^b}A_{b}(\Lambda^{-1}x,N).\qquad
\end{eqnarray}
Recall the electromagnetic field operator (\ref{e-m1})
\begin{eqnarray}
&&\hspace{-6pt}F_{ab}(x,1)
\nonumber\\
&&=
\int d\Gamma({\bf k})
\left(
k_a({\bf k})
g_{b}{^{\bf{a}}}({\bf k})
-
k_{b}({\bf k})
g_{a}{^{\bf{a}}}({\bf k})
\right)
a_{\bf{a}}({\bf k},1)
e^{-ik\cdot x}
+{\rm H.c.}
\nonumber
\end{eqnarray}
The electromagnetic field operator transforms under Lorentz transformation (\ref{WU}) like a tensor
\begin{eqnarray}
U(\Lambda,0,1)^{\dagger} F_{ab}(x,1) U(\Lambda,0,1)
~=~
\Lambda_{a}{^{c}}
\Lambda_{b}{^{d}}
F_{cd}(\Lambda^{-1}x,1).
\end{eqnarray}
{\bf Proof}:
\begin{eqnarray}
&&\hspace{-6pt}U(\Lambda,0,1)^{\dagger} F_{ab}(x,1) U(\Lambda,0,1)
\nonumber\\
&=&
2\int d\Gamma({\bf k})
k_{[a}({\bf k})
g_{b]}{^{\bf{a}}}({\bf k})
R_{\bf{a}}{^{\bf{b}}}(\Lambda,{\bf k})
a_{\bf{b}}({\bf{\Lambda^{-1}k}},1)
e^{-ik\cdot x}
+{\rm H.c.}
\nonumber\\
&=&
2\int d\Gamma({\bf k})
\Lambda k_{[a}({\bf k})
\Lambda g_{b]}{^{\bf{a}}}({\bf k})
a_{\bf{b}}({\bf{\Lambda^{-1}k}},1)
e^{-ik\cdot x}
+{\rm H.c.}
\nonumber\\
&=&
2\int d\Gamma({\bf k})
\Lambda k_{[a}({\bf \Lambda k})
\Lambda g_{b]}{^{\bf{a}}}({\bf \Lambda k})
a_{\bf{b}}({\bf{k}},1)
e^{-i\Lambda k\cdot x}
+{\rm H.c.}
\nonumber\\
&=&
2
\Lambda_{a}{^{c}}
\Lambda_{b}{^{d}}
\int d\Gamma({\bf k})
k_{[c}({\bf k})
g_{d]}^{~~{\bf a}}({\bf k})
a_{\bf{a}}({\bf k},1)
e^{-i k\cdot \Lambda^{-1}x}+{\rm H.c.}~
\nonumber\\
&=&
\Lambda_{a}{^{c}}
\Lambda_{b}{^{d}}
F_{cd}(\Lambda^{-1}x,1)
	\end{eqnarray}
The same can be shown for any $N$-oscillator representation.
\end{small}
%
%
%
%
%



\end{document}